
 %
 \catcode`@=11 
 %
 %
 %
 %
 
 \font\fourteenrm=cmr10 scaled\magstep2
 \font\twelverm=cmr10 scaled\magstep1
 \font\ninerm=cmr9            \font\sixrm=cmr6

 \font\fourteenbf=cmbx10 scaled\magstep2
 \font\twelvebf=cmbx10 scaled\magstep1
 \font\ninebf=cmbx9            \font\sixbf=cmbx6
 \font\seventeeni=cmmi10 scaled\magstep3     \skewchar\seventeeni='177
 \font\fourteeni=cmmi10 scaled\magstep2      \skewchar\fourteeni='177
 \font\twelvei=cmmi10 scaled\magstep1        \skewchar\twelvei='177
 \font\ninei=cmmi9                           \skewchar\ninei='177
 \font\sixi=cmmi6                            \skewchar\sixi='177
 \font\seventeensy=cmsy10 scaled\magstep3    \skewchar\seventeensy='60
 \font\fourteensy=cmsy10 scaled\magstep2     \skewchar\fourteensy='60
 \font\twelvesy=cmsy10 scaled\magstep1       \skewchar\twelvesy='60
 \font\ninesy=cmsy9                          \skewchar\ninesy='60
 \font\sixsy=cmsy6                           \skewchar\sixsy='60

 \font\fourteenex=cmex10 scaled\magstep2
 \font\twelveex=cmex10 scaled\magstep1

 \font\fourteensl=cmti10 scaled\magstep2
 \font\twelvesl=cmti10 scaled\magstep1
 \font\ninesl=cmti9

 \font\fourteenit=cmti10 scaled\magstep2
 \font\twelveit=cmti10 scaled\magstep1
 \font\twelvett=cmtt10 scaled\magstep1
 \font\twelvecp=cmcsc10 scaled\magstep1
 \font\tencp=cmcsc10
 \newfam\cpfam
 \font\tenfib=cmr10      
 \newcount\f@ntkey            \f@ntkey=0
 \def\samef@nt{\relax \ifcase\f@ntkey \rm \or\oldstyle \or\or
          \or\it \or\sl \or\bf \or\tt \or\caps \fi }
 \def\fourteenpoint{\relax
     \textfont0=\fourteenrm          \scriptfont0=\tenrm
     \scriptscriptfont0=\sevenrm
      \def\rm{\fam0 \fourteenrm \f@ntkey=0 }\relax
     \textfont1=\fourteeni           \scriptfont1=\teni
     \scriptscriptfont1=\seveni
      \def\oldstyle{\fam1 \fourteeni\f@ntkey=1 }\relax
     \textfont2=\fourteensy          \scriptfont2=\tensy
     \scriptscriptfont2=\sevensy
     \textfont3=\fourteenex     \scriptfont3=\fourteenex
     \scriptscriptfont3=\fourteenex
     \def\it{\fam\itfam \fourteenit\f@ntkey=4 }\textfont\itfam=\fourteenit
     \def\sl{\fam\slfam \fourteensl\f@ntkey=5 }\textfont\slfam=\fourteensl
     \scriptfont\slfam=\tensl
     \def\bf{\fam\bffam \fourteenbf\f@ntkey=6 }\textfont\bffam=\fourteenbf
     \scriptfont\bffam=\tenbf     \scriptscriptfont\bffam=\sevenbf
     \def\tt{\fam\ttfam \twelvett \f@ntkey=7 }\textfont\ttfam=\twelvett
     \h@big=11.9\p@{} \h@Big=16.1\p@{} \h@bigg=20.3\p@{} \h@Bigg=24.5\p@{}
     \def\caps{\fam\cpfam \twelvecp \f@ntkey=8 }\textfont\cpfam=\twelvecp
     \setbox\strutbox=\hbox{\vrule height 12pt depth 5pt width\z@}
     \samef@nt}
 \def\twelvepoint{\relax
     \textfont0=\twelverm          \scriptfont0=\ninerm
     \scriptscriptfont0=\sixrm
      \def\rm{\fam0 \twelverm \f@ntkey=0 }\relax
     \textfont1=\twelvei           \scriptfont1=\ninei
     \scriptscriptfont1=\sixi
      \def\oldstyle{\fam1 \twelvei\f@ntkey=1 }\relax
     \textfont2=\twelvesy          \scriptfont2=\ninesy
     \scriptscriptfont2=\sixsy
     \textfont3=\twelveex          \scriptfont3=\twelveex
     \scriptscriptfont3=\twelveex
     \def\it{\fam\itfam \twelveit \f@ntkey=4 }\textfont\itfam=\twelveit
     \def\sl{\fam\slfam \twelvesl \f@ntkey=5 }\textfont\slfam=\twelvesl
     \scriptfont\slfam=\ninesl
     \def\bf{\fam\bffam \twelvebf \f@ntkey=6 }\textfont\bffam=\twelvebf
     \scriptfont\bffam=\ninebf     \scriptscriptfont\bffam=\sixbf
     \def\tt{\fam\ttfam \twelvett \f@ntkey=7 }\textfont\ttfam=\twelvett
     \h@big=10.2\p@{}
     \h@Big=13.8\p@{}
     \h@bigg=17.4\p@{}
     \h@Bigg=21.0\p@{}
     \def\caps{\fam\cpfam \twelvecp \f@ntkey=8 }\textfont\cpfam=\twelvecp
     \setbox\strutbox=\hbox{\vrule height 10pt depth 4pt width\z@}
     \samef@nt}
 \def\tenpoint{\relax
     \textfont0=\tenrm          \scriptfont0=\sevenrm
     \scriptscriptfont0=\fiverm
     \def\rm{\fam0 \tenrm \f@ntkey=0 }\relax
     \textfont1=\teni           \scriptfont1=\seveni
     \scriptscriptfont1=\fivei
     \def\oldstyle{\fam1 \teni \f@ntkey=1 }\relax
     \textfont2=\tensy          \scriptfont2=\sevensy
     \scriptscriptfont2=\fivesy
     \textfont3=\tenex          \scriptfont3=\tenex
     \scriptscriptfont3=\tenex
     \def\it{\fam\itfam \tenit \f@ntkey=4 }\textfont\itfam=\tenit
     \def\sl{\fam\slfam \tensl \f@ntkey=5 }\textfont\slfam=\tensl
     \def\bf{\fam\bffam \tenbf \f@ntkey=6 }\textfont\bffam=\tenbf
     \scriptfont\bffam=\sevenbf     \scriptscriptfont\bffam=\fivebf
     \def\tt{\fam\ttfam \tentt \f@ntkey=7 }\textfont\ttfam=\tentt
     \def\caps{\fam\cpfam \tencp \f@ntkey=8 }\textfont\cpfam=\tencp
     \setbox\strutbox=\hbox{\vrule height 8.5pt depth 3.5pt width\z@}
     \samef@nt}
 %
 %
 %
 %
 \newdimen\h@big  \h@big=8.5\p@
 \newdimen\h@Big  \h@Big=11.5\p@
 \newdimen\h@bigg  \h@bigg=14.5\p@
 \newdimen\h@Bigg  \h@Bigg=17.5\p@
 \def\big#1{{\hbox{$\left#1\vbox to\h@big{}\right.\n@space$}}}
 \def\Big#1{{\hbox{$\left#1\vbox to\h@Big{}\right.\n@space$}}}
 \def\bigg#1{{\hbox{$\left#1\vbox to\h@bigg{}\right.\n@space$}}}
 \def\Bigg#1{{\hbox{$\left#1\vbox to\h@Bigg{}\right.\n@space$}}}
 %
 %
 %
 \normalbaselineskip = 20pt plus 0.2pt minus 0.1pt
 \normallineskip = 1.5pt plus 0.1pt minus 0.1pt
 \normallineskiplimit = 1.5pt
 \newskip\normaldisplayskip
 \normaldisplayskip = 20pt plus 5pt minus 10pt
 \newskip\normaldispshortskip
 \normaldispshortskip = 6pt plus 5pt
 \newskip\normalparskip
 \normalparskip = 6pt plus 2pt minus 1pt
 \newskip\skipregister
 \skipregister = 5pt plus 2pt minus 1.5pt
 \newif\ifsingl@    \newif\ifdoubl@
 \newif\iftwelv@    \twelv@true
 \def\singlespace{\singl@true\doubl@false\spaces@t}
 \def\doublespace{\singl@false\doubl@true\spaces@t}
 \def\normalspace{\singl@false\doubl@false\spaces@t}
 \def\Tenpoint{\tenpoint\twelv@false\spaces@t}
 \def\Twelvepoint{\twelvepoint\twelv@true\spaces@t}
 \def\spaces@t{\relax%
  \iftwelv@ \ifsingl@\subspaces@t3:4;\else\subspaces@t1:1;\fi%
  \else \ifsingl@\subspaces@t3:5;\else\subspaces@t4:5;\fi \fi%
  \ifdoubl@ \multiply\baselineskip by 5%
  \divide\baselineskip by 4 \fi \unskip}
 \def\subspaces@t#1:#2;{
       \baselineskip = \normalbaselineskip
       \multiply\baselineskip by #1 \divide\baselineskip by #2
       \lineskip = \normallineskip
       \multiply\lineskip by #1 \divide\lineskip by #2
       \lineskiplimit = \normallineskiplimit
       \multiply\lineskiplimit by #1 \divide\lineskiplimit by #2
       \parskip = \normalparskip
       \multiply\parskip by #1 \divide\parskip by #2
       \abovedisplayskip = \normaldisplayskip
       \multiply\abovedisplayskip by #1 \divide\abovedisplayskip by #2
       \belowdisplayskip = \abovedisplayskip
       \abovedisplayshortskip = \normaldispshortskip
       \multiply\abovedisplayshortskip by #1
         \divide\abovedisplayshortskip by #2
       \belowdisplayshortskip = \abovedisplayshortskip
       \advance\belowdisplayshortskip by \belowdisplayskip
       \divide\belowdisplayshortskip by 2
       \smallskipamount = \skipregister
       \multiply\smallskipamount by #1 \divide\smallskipamount by #2
       \medskipamount = \smallskipamount \multiply\medskipamount by 2
       \bigskipamount = \smallskipamount \multiply\bigskipamount by 4 }
 \def\normalbaselines{ \baselineskip=\normalbaselineskip
    \lineskip=\normallineskip \lineskiplimit=\normallineskip
    \iftwelv@\else \multiply\baselineskip by 4 \divide\baselineskip by 5
      \multiply\lineskiplimit by 4 \divide\lineskiplimit by 5
      \multiply\lineskip by 4 \divide\lineskip by 5 \fi }
 \Twelvepoint  
 \interlinepenalty=50
 \interfootnotelinepenalty=5000
 \predisplaypenalty=9000
 \postdisplaypenalty=500
 \hfuzz=1pt
 \vfuzz=0.2pt
 %
 %
 %
 \def\pagecontents{
    \ifvoid\topins\else\unvbox\topins\vskip\skip\topins\fi
    \dimen@ = \dp255 \unvbox255
    \ifvoid\footins\else\vskip\skip\footins\footrule\unvbox\footins\fi
    \ifr@ggedbottom \kern-\dimen@ \vfil \fi }
 \def\makeheadline{\vbox to 0pt{ \skip@=\topskip
       \advance\skip@ by -12pt \advance\skip@ by -2\normalbaselineskip
       \vskip\skip@ \line{\vbox to 12pt{}\the\headline} \vss
       }\nointerlineskip}
 \def\makefootline{\baselineskip = 1.5\normalbaselineskip
                  \line{\the\footline}}
 \newif\iffrontpage
 \newif\ifletterstyle
 \newif\ifp@genum
 \def\nopagenumbers{\p@genumfalse}
 \def\pagenumbers{\p@genumtrue}
 \pagenumbers
 \newtoks\paperheadline
 \newtoks\letterheadline
 \newtoks\letterfrontheadline
 \newtoks\lettermainheadline
 \newtoks\paperfootline
 \newtoks\letterfootline
 \newtoks\date
 \footline={\ifletterstyle\the\letterfootline\else\the\paperfootline\fi}
 \paperfootline={\hss\iffrontpage\else\ifp@genum\tenrm\folio\hss\fi\fi}
 \letterfootline={\hfil}
 \headline={\ifletterstyle\the\letterheadline\else\the\paperheadline\fi}
 \paperheadline={\hfil}
 \letterheadline{\iffrontpage\the\letterfrontheadline
      \else\the\lettermainheadline\fi}
 \lettermainheadline={\rm\ifp@genum page \ \folio\fi\hfil\the\date}
 \def\monthname{\relax\ifcase\month 0/\or January\or February\or
    March\or April\or May\or June\or July\or August\or September\or
    October\or November\or December\else\number\month/\fi}
 \date={\monthname\ \number\day, \number\year}
 \countdef\pagenumber=1  \pagenumber=1
 \def\advancepageno{\global\advance\pageno by 1
    \ifnum\pagenumber<0 \global\advance\pagenumber by -1
     \else\global\advance\pagenumber by 1 \fi \global\frontpagefalse }
 \def\folio{\ifnum\pagenumber<0 \romannumeral-\pagenumber
            \else \number\pagenumber \fi }
 \def\footrule{\dimen@=\prevdepth\nointerlineskip
    \vbox to 0pt{\vskip -0.25\baselineskip \hrule width 0.35\hsize \vss}
    \prevdepth=\dimen@ }
 \newtoks\foottokens
 \foottokens={\Tenpoint\singlespace}
 \newdimen\footindent
 \footindent=24pt
 \def\vfootnote#1{\insert\footins\bgroup  \the\foottokens
    \interlinepenalty=\interfootnotelinepenalty \floatingpenalty=20000
    \splittopskip=\ht\strutbox \boxmaxdepth=\dp\strutbox
    \leftskip=\footindent \rightskip=\z@skip
    \parindent=0.5\footindent \parfillskip=0pt plus 1fil
    \spaceskip=\z@skip \xspaceskip=\z@skip
    \Textindent{$ #1 $}\footstrut\futurelet\next\fo@t}
 \def\Textindent#1{\noindent\llap{#1\enspace}\ignorespaces}
 \def\footnote#1{\attach{#1}\vfootnote{#1}}

 \let\footsymbol=\star
 \newcount\lastf@@t           \lastf@@t=-1
 \newcount\footsymbolcount    \footsymbolcount=0
 \newif\ifPhysRev
 \def\footsymbolgen{\relax \ifPhysRev \iffrontpage \NPsymbolgen\else
       \PRsymbolgen\fi \else \NPsymbolgen\fi
    \global\lastf@@t=\pageno \footsymbol }
 \def\NPsymbolgen{\ifnum\footsymbolcount<0 \global\footsymbolcount=0\fi
    {\iffrontpage \else \advance\lastf@@t by 1 \fi
     \ifnum\lastf@@t<\pageno \global\footsymbolcount=0
      \else \global\advance\footsymbolcount by 1 \fi }
    \ifcase\footsymbolcount \fd@f\star\or \fd@f\dagger\or \fd@f\ast\or
     \fd@f\ddagger\or \fd@f\natural\or \fd@f\diamond\or \fd@f\bullet\or
     \fd@f\nabla\else \fd@f\dagger\global\footsymbolcount=0 \fi }
 \def\fd@f#1{\xdef\footsymbol{#1}}
 \def\PRsymbolgen{\ifnum\footsymbolcount>0 \global\footsymbolcount=0\fi
       \global\advance\footsymbolcount by -1
       \xdef\footsymbol{\sharp\number-\footsymbolcount} }
 \def\space@ver#1{\let\@sf=\empty \ifmmode #1\else \ifhmode
    \edef\@sf{\spacefactor=\the\spacefactor}\unskip${}#1$\relax\fi\fi}
 \def\attach#1{\space@ver{\strut^{\mkern 2mu #1} }\@sf\ }
 %
 %
 %
 \newcount\chapternumber      \chapternumber=0
 \newcount\sectionnumber      \sectionnumber=0
 \newcount\equanumber         \equanumber=0
 \let\chapterlabel=0
 \newtoks\chapterstyle        \chapterstyle={\Number}
 \newskip\chapterskip         \chapterskip=\bigskipamount
 \newskip\sectionskip         \sectionskip=\medskipamount
 \newskip\headskip            \headskip=8pt plus 3pt minus 3pt
 \newdimen\chapterminspace    \chapterminspace=15pc
 \newdimen\sectionminspace    \sectionminspace=10pc
 \newdimen\referenceminspace  \referenceminspace=25pc
 \def\chapterreset{\global\advance\chapternumber by 1
    \ifnum\the\equanumber<0 \else\global\equanumber=0\fi
    \sectionnumber=0 \makel@bel}
 \def\makel@bel{\xdef\chapterlabel{%
 \the\chapterstyle{\the\chapternumber}.}}
 \def\sectionlabel{\number\sectionnumber \quad }
 \def\alphabetic#1{\count255='140 \advance\count255 by #1\char\count255}
 \def\Alphabetic#1{\count255='100 \advance\count255 by #1\char\count255}
 \def\Roman#1{\uppercase\expandafter{\romannumeral #1}}
 \def\roman#1{\romannumeral #1}
 \def\Number#1{\number #1}
 \def\unnumberedchapters{\let\makel@bel=\relax \let\chapterlabel=\relax
 \let\sectionlabel=\relax \equanumber=-1 }
 \def\titlestyle#1{\par\begingroup \interlinepenalty=9999
      \leftskip=0.02\hsize plus 0.23\hsize minus 0.02\hsize
      \rightskip=\leftskip \parfillskip=0pt
      \hyphenpenalty=9000 \exhyphenpenalty=9000
      \tolerance=9999 \pretolerance=9000
      \spaceskip=0.333em \xspaceskip=0.5em
      \iftwelv@\fourteenpoint\else\twelvepoint\fi
    \noindent #1\par\endgroup }
 \def\spacecheck#1{\dimen@=\pagegoal\advance\dimen@ by -\pagetotal
    \ifdim\dimen@<#1 \ifdim\dimen@>0pt \vfil\break \fi\fi}
 \def\chapter#1{\par \penalty-300 \vskip\chapterskip
    \spacecheck\chapterminspace
    \chapterreset \titlestyle{\chapterlabel \ #1}
    \nobreak\vskip\headskip \penalty 30000
    \wlog{\string\chapter\ \chapterlabel} }

 \def\section#1{\par \ifnum\the\lastpenalty=30000\else
    \penalty-200\vskip\sectionskip \spacecheck\sectionminspace\fi
    \wlog{\string\section\ \chapterlabel \the\sectionnumber}
    \global\advance\sectionnumber by 1  \noindent
    {\caps\enspace\chapterlabel \sectionlabel #1}\par
    \nobreak\vskip\headskip \penalty 30000 }
 \def\subsection#1{\par
    \ifnum\the\lastpenalty=30000\else \penalty-100\smallskip \fi
    \noindent\undertext{#1}\enspace \vadjust{\penalty5000}}

 \def\undertext#1{\vtop{\hbox{#1}\kern 1pt \hrule}}
 \def\ack{\par\penalty-100\medskip \spacecheck\sectionminspace
    \line{\fourteenrm\hfil ACKNOWLEDGEMENTS\hfil}\nobreak\vskip\headskip }
 \def\APPENDIX#1#2{\par\penalty-300\vskip\chapterskip
    \spacecheck\chapterminspace \chapterreset \xdef\chapterlabel{#1}
    \titlestyle{APPENDIX #2} \nobreak\vskip\headskip \penalty 30000
    \wlog{\string\Appendix\ \chapterlabel} }
 \def\Appendix#1{\APPENDIX{#1}{#1}}
 \def\appendix{\APPENDIX{A}{}}
 %
 %
 %
 \def\eqname#1{\relax \ifnum\the\equanumber<0
      \xdef#1{{\rm(\number-\equanumber)}}\global\advance\equanumber by -1
     \else \global\advance\equanumber by 1
       \xdef#1{{\rm(\chapterlabel \number\equanumber)}} \fi}

 \def\eqn#1{\eqno\eqname{#1}#1}

 \def\eqinsert#1{\noalign{\dimen@=\prevdepth \nointerlineskip
    \setbox0=\hbox to\displaywidth{\hfil #1}
    \vbox to 0pt{\vss\hbox{$\!\box0\!$}\kern-0.5\baselineskip}
    \prevdepth=\dimen@}}
 \def\eqnalign#1{\eqname{#1}#1}
 %
 
 %
 %
 \def\GENITEM#1;#2{\par \hangafter=0 \hangindent=#1
     \Textindent{$ #2 $}\ignorespaces}
 \outer\def\newitem#1=#2;{\gdef#1{\GENITEM #2;}}
 \newdimen\itemsize                \itemsize=30pt
 \newitem\item=1\itemsize;
 \newitem\sitem=1.75\itemsize;     
 \newitem\ssitem=2.5\itemsize;     
 \outer\def\newlist#1=#2&#3&#4;{\toks0={#2}\toks1={#3}%
    \count255=\escapechar \escapechar=-1
    \alloc@0\list\countdef\insc@unt\listcount     \listcount=0
    \edef#1{\par
       \countdef\listcount=\the\allocationnumber
       \advance\listcount by 1
       \hangafter=0 \hangindent=#4
       \Textindent{\the\toks0{\listcount}\the\toks1}}
    \expandafter\expandafter\expandafter
     \edef\c@t#1{begin}{\par
       \countdef\listcount=\the\allocationnumber \listcount=1
       \hangafter=0 \hangindent=#4
       \Textindent{\the\toks0{\listcount}\the\toks1}}
    \expandafter\expandafter\expandafter
     \edef\c@t#1{con}{\par \hangafter=0 \hangindent=#4 \noindent}
    \escapechar=\count255}
 \def\c@t#1#2{\csname\string#1#2\endcsname}
 \newlist\point=\Number&.&1.0\itemsize;
 \newlist\subpoint=(\alphabetic&)&1.75\itemsize;
 \newlist\subsubpoint=(\roman&)&2.5\itemsize;
 \newlist\cpoint=\Roman&.&1.0\itemsize;
 %

 %
 %
 %
 \newcount\referencecount     \referencecount=0
 \newif\ifreferenceopen       \newwrite\referencewrite
 \newtoks\rw@toks
 \def\NPrefmark#1{\attach{\scriptscriptstyle [ #1 ] }}
 \let\PRrefmark=\attach
 \def\CErefmark#1{\attach{\scriptstyle  #1 ) }}
 \def\refmark#1{\relax\ifPhysRev\PRrefmark{#1}\else\NPrefmark{#1}\fi}
 \def\crefmark#1{\relax\CErefmark{#1}}
 \def\refend{\refmark{\number\referencecount}}
 \newcount\lastrefsbegincount \lastrefsbegincount=0
 \def\refsend{\refmark{\count255=\referencecount
    \advance\count255 by-\lastrefsbegincount
    \ifcase\count255 \number\referencecount
    \or \number\lastrefsbegincount,\number\referencecount
    \else \number\lastrefsbegincount-\number\referencecount \fi}}
 \def\crefsend{\crefmark{\count255=\referencecount
    \advance\count255 by-\lastrefsbegincount
    \ifcase\count255 \number\referencecount
    \or \number\lastrefsbegincount,\number\referencecount
    \else \number\lastrefsbegincount-\number\referencecount \fi}}
 \def\refch@ck{\chardef\rw@write=\referencewrite
    \ifreferenceopen \else \referenceopentrue
    \immediate\openout\referencewrite=referenc.texauxil \fi}
 %
 {\catcode`\^^M=\active 
   \gdef\obeyendofline{\catcode`\^^M\active \let^^M\ }}%
 %
 {\catcode`\^^M=\active 
   \gdef\ignoreendofline{\catcode`\^^M=5}}
 {\obeyendofline\gdef\rw@start#1{\def\t@st{#1} \ifx\t@st\blankend%
 \endgroup \@sf \relax \else \ifx\t@st\bl@nkend \endgroup \@sf \relax%
 \else \rw@begin#1
 \backtotext
 \fi \fi } }
 {\obeyendofline\gdef\rw@begin#1
 {\def\n@xt{#1}\rw@toks={#1}\relax%
 \rw@next}}
 \def\blankend{}
 {\obeylines\gdef\bl@nkend{
 }}
 \newif\iffirstrefline  \firstreflinetrue
 \def\rwr@teswitch{\ifx\n@xt\blankend \let\n@xt=\rw@begin %
  \else\iffirstrefline \global\firstreflinefalse%
 \immediate\write\rw@write{\noexpand\obeyendofline \the\rw@toks}%
 \let\n@xt=\rw@begin%
       \else\ifx\n@xt\rw@@d \def\n@xt{\immediate\write\rw@write{%
         \noexpand\ignoreendofline}\endgroup \@sf}%
              \else \immediate\write\rw@write{\the\rw@toks}%
              \let\n@xt=\rw@begin\fi\fi \fi}
 \def\rw@next{\rwr@teswitch\n@xt}
 \def\rw@@d{\backtotext} \let\rw@end=\relax
 \let\backtotext=\relax

 \newdimen\refindent     \refindent=30pt
 \def\refitem#1{\par \hangafter=0 \hangindent=\refindent \Textindent{#1}}
 \def\REFNUM#1{\space@ver{}\refch@ck \firstreflinetrue%
  \global\advance\referencecount by 1 \xdef#1{\the\referencecount}}
 \def\refnum#1{\space@ver{}\refch@ck \firstreflinetrue%
  \global\advance\referencecount by 1 \xdef#1{\the\referencecount}\refend}

 \def\REF#1{\REFNUM#1%
  \immediate\write\referencewrite{%
  \noexpand\refitem{#1.}}%
 \begingroup\obeyendofline\rw@start}
 \def\ref{\refnum\?%
  \immediate\write\referencewrite{\noexpand\refitem{\?.}}%
 \begingroup\obeyendofline\rw@start}
 \def\Ref#1{\refnum#1%
  \immediate\write\referencewrite{\noexpand\refitem{#1.}}%
 \begingroup\obeyendofline\rw@start}
 \def\REFS#1{\REFNUM#1\global\lastrefsbegincount=\referencecount
 \immediate\write\referencewrite{\noexpand\refitem{#1.}}%
 \begingroup\obeyendofline\rw@start}
 \newcount\figurecount     \figurecount=0
 \newif\iffigureopen       \newwrite\figurewrite
 \def\figch@ck{\chardef\rw@write=\figurewrite \iffigureopen\else
    \immediate\openout\figurewrite=figures.texauxil
    \figureopentrue\fi}
 \def\FIGNUM#1{\space@ver{}\figch@ck \firstreflinetrue%
  \global\advance\figurecount by 1 \xdef#1{\the\figurecount}}
 \def\FIG#1{\FIGNUM#1
    \immediate\write\figurewrite{\noexpand\refitem{#1.}}%
    \begingroup\obeyendofline\rw@start}
 \def\fig{\FIGNUM\? fig.
 \immediate\write\figurewrite{\noexpand\refitem{\?.}}%
 \begingroup\obeyendofline\rw@start}
 \def\figure{\FIGNUM\? figure
    \immediate\write\figurewrite{\noexpand\refitem{\?.}}%
    \begingroup\obeyendofline\rw@start}
 \def\Fig{\FIGNUM\? Fig.
 \immediate\write\figurewrite{\noexpand\refitem{\?.}}%
 \begingroup\obeyendofline\rw@start}
 \def\Figure{\FIGNUM\? Figure
 \immediate\write\figurewrite{\noexpand\refitem{\?.}}%
 \begingroup\obeyendofline\rw@start}
 \newcount\tablecount     \tablecount=0
 \newif\iftableopen       \newwrite\tablewrite
 \def\tabch@ck{\chardef\rw@write=\tablewrite \iftableopen\else
    \immediate\openout\tablewrite=tables.texauxil
    \tableopentrue\fi}
 \def\TABNUM#1{\space@ver{}\tabch@ck \firstreflinetrue%
  \global\advance\tablecount by 1 \xdef#1{\the\tablecount}}
 \def\TABLE#1{\TABNUM#1
    \immediate\write\tablewrite{\noexpand\refitem{#1.}}%
    \begingroup\obeyendofline\rw@start}
 \def\Table{\TABNUM\? Table
 \immediate\write\tablewrite{\noexpand\refitem{\?.}}%
 \begingroup\obeyendofline\rw@start}
 %
 
 %
 %
 %
 \def\masterreset{\global\pagenumber=1 \global\chapternumber=0
    \ifnum\the\equanumber<0\else \global\equanumber=0\fi
    \global\sectionnumber=0
    \global\referencecount=0 \global\figurecount=0 \global\tablecount=0 }
 \def\FRONTPAGE{\ifvoid255\else\vfill\penalty-2000\fi
       \masterreset\global\frontpagetrue
       \global\lastf@@t=0 \global\footsymbolcount=0}

 \def\paperstyle{\letterstylefalse\normalspace\papersize}
 \def\letterstyle{\letterstyletrue\singlespace\lettersize}
 \def\papersize{\hsize=38pc\vsize=56pc\hoffset=1pc\voffset=0pc
                \skip\footins=\bigskipamount}
 \def\lettersize{\hsize=6.5in\vsize=8.5in\hoffset=0in\voffset=1in
    \skip\footins=\smallskipamount \multiply\skip\footins by 3 }
 \paperstyle   
 %
 %
 \def\MEMO{\letterstyle\FRONTPAGE \letterfrontheadline={\hfil}
     \line{\quad\fourteenrm DURHAM MEMORANDUM\hfil\twelverm\the\date\quad}
     \medskip \memod@f}

 \def\memit@m#1{\smallskip \hangafter=0 \hangindent=1in
       \Textindent{\caps #1}}
 \def\memod@f{\xdef\to{\memit@m{To:}}\xdef\from{\memit@m{From:}}%
      \xdef\topic{\memit@m{Topic:}}\xdef\subject{\memit@m{Subject:}}%
      \xdef\rule{\bigskip\hrule height 1pt\bigskip}}
 \memod@f

 \newskip\lettertopfil
 \lettertopfil = 0pt plus 1.5in minus 0pt
 \newskip\letterbottomfil
 \letterbottomfil = 0pt plus 2.3in minus 0pt
 \newskip\spskip \setbox0\hbox{\ } \spskip=-1\wd0
 \def\addressee#1{\medskip\rightline{{\tenfib\the\date}\hskip 30pt} \bigskip
    \vskip\lettertopfil
    \ialign to\hsize{\strut ##\hfil\tabskip 0pt plus \hsize \cr #1\crcr}
    \medskip\noindent\hskip\spskip}
 \newskip\signatureskip       \signatureskip=40pt
 \def\signed#1{\par \penalty 9000 \bigskip \dt@pfalse
   \everycr={\noalign{\ifdt@p\vskip\signatureskip\global\dt@pfalse\fi}}
   \setbox0=\vbox{\singlespace \halign{\tabskip 0pt \strut ##\hfil\cr
    \noalign{\global\dt@ptrue}#1\crcr}}
   \line{\hskip 0.5\hsize minus 0.5\hsize \box0\hfil} \medskip }

 \def\endletter{\ifnum\pagenumber=1 \vskip\letterbottomfil\supereject
 \else \vfil\supereject \fi}
 \newbox\letterb@x
 \def\lettertext{\par\unvcopy\letterb@x\par}
 \def\multiletter{\setbox\letterb@x=\vbox\bgroup
       \everypar{\vrule height 1\baselineskip depth 0pt width 0pt }
       \singlespace \topskip=\baselineskip }
 \def\letterend{\par\egroup}
 %
 %
 %
 \newskip\frontpageskip
 \newtoks\pubtype
 \newtoks\Pubnum
 \newtoks\pubnum
 \newif\ifp@bblock  \p@bblocktrue
 \def\PH@SR@V{\doubl@true \baselineskip=24.1pt plus 0.2pt minus 0.1pt
              \parskip= 3pt plus 2pt minus 1pt }
 \def\PHYSREV{\paperstyle\PhysRevtrue\PH@SR@V}
 \def\titlepage{\FRONTPAGE\paperstyle\ifPhysRev\PH@SR@V\fi
   \ifp@bblock\p@bblock\fi}
 \def\nopubblock{\p@bblockfalse}
 \def\endpage{\vfil\break}
 \frontpageskip=1\medskipamount plus .5fil
 \pubtype={\tensl Preliminary Version}
\Pubnum={89--17P}
 \pubnum={0000}
 \def\p@bblock{\begingroup \tabskip=\hsize minus \hsize
    \baselineskip=1.5\ht\strutbox \topspace-2\baselineskip
    \halign to\hsize{\strut ##\hfil\tabskip=0pt\crcr
    \the\Pubnum\cr \the\date\cr}\endgroup}
 \def\title#1{\vskip\frontpageskip \titlestyle{#1} \vskip\headskip }
 \def\author#1{\vskip\frontpageskip\titlestyle{\twelvecp #1}\nobreak}
 \def\andauthor{\vskip\frontpageskip\centerline{and}\author}

 \def\address#1{\par\kern 5pt\titlestyle{\twelveit #1}}
 \def\andaddress{\par\kern 5pt \centerline{\sl and} \address}

 \def\abstract{\vskip\frontpageskip\centerline{\fourteenrm ABSTRACT}
               \vskip\headskip }

 %
 %
 %

 \def\\{\relax\ifmmode\backslash\else$\backslash$\fi}
 \def\globaleqnumbers{\relax\ifnum\the\equanumber<0%
 \else\global\equanumber=-1\fi}

\def\journal#1&#2(#3){\unskip, \sl #1 \bf #2 \rm (19#3) }

 \def\topspace{\hrule height 0pt depth 0pt \vskip}

 \let\int=\intop         
 \def\prop{\mathrel{{\mathchoice{\pr@p\scriptstyle}{\pr@p\scriptstyle}{
                 \pr@p\scriptscriptstyle}{\pr@p\scriptscriptstyle} }}}
 \def\pr@p#1{\setbox0=\hbox{$\cal #1 \char'103$}
    \hbox{$\cal #1 \char'117$\kern-.4\wd0\box0}}
 \def\lsim{\mathrel{\mathpalette\@versim<}}
 \def\gsim{\mathrel{\mathpalette\@versim>}}
 \def\@versim#1#2{\lower0.2ex\vbox{\baselineskip\z@skip\lineskip\z@skip
   \lineskiplimit\z@\ialign{$\m@th#1\hfil##\hfil$\crcr#2\crcr\sim\crcr}}}
 \def\leftrightarrowfill{$\m@th \mathord- \mkern-6mu
         \cleaders\hbox{$\mkern-2mu \mathord- \mkern-2mu$}\hfil
         \mkern-6mu \mathord\leftrightarrow$}
 \def\lrover#1{\vbox{\ialign{##\crcr
         \leftrightarrowfill\crcr\noalign{\kern-1pt\nointerlineskip}
         $\hfil\displaystyle{#1}\hfil$\crcr}}}
 %
 %
 %
 \let\sec@nt=\sec
 \def\sec{\relax\ifmmode\let\n@xt=\sec@nt\else\let\n@xt\section\fi\n@xt}
 \def\obsolete#1{\message{Macro \string #1 is obsolete.}}
 \def\firstsec#1{\obsolete\firstsec \section{#1}}
 \def\firstsubsec#1{\obsolete\firstsubsec \subsection{#1}}
 \def\thispage#1{\obsolete\thispage \global\pagenumber=#1\frontpagefalse}
 \def\thischapter#1{\obsolete\thischapter \global\chapternumber=#1}
 \def\nextequation#1{\obsolete\nextequation \global\equanumber=#1
    \ifnum\the\equanumber>0 \global\advance\equanumber by 1 \fi}
 \def\BOXITEM{\afterassigment\B@XITEM\setbox0=}
 \def\B@XITEM{\par\hangindent\wd0 \noindent\box0 }
 %
 
 %
 \catcode`@=12 
 \message{ by V.K.}
%
\def\NPrefmark#1{ [#1] }
%
\hsize=38pc
\vsize=50pc
\hoffset=-2pc
\voffset=-2pc

\def\SB{Kac, V.G.: Infinite Dimensional Lie Algebras, $2^{nd}$ edition.
Cambridge University Press 1985}

\def\SC{Drinfel'd, V.G., Sokolov, V.V.: Lie Algebras and Equations of the
 Korteweg-de Vries
Type. Jour.Sov.Math. {\bf 30} (1985) 1975;
Equations of Korteweg-De Vries Type and
Simple Lie Algebras. Soviet.Math.Dokl. {\bf23} (1981) 457}

\def\SE{Bakas, I., Depireux, D.A.: The Origins of Gauge Symmetries in
Integrable
Systems of
KdV Type. Univ. of Maryland preprint, UMD-PP91-111(1990);
A Fractional KdV Hierarchy, Univ. of Maryland
preprint,
UMD-PP91-168(1990)}

\def\SG{Bershadsky, M.: Conformal Field Theories via Hamiltonian Reduction,IAS
pre- print, IASSNS-HEP-90/44\hfil\break
Polyakov, A.: Gauge Transformations and Diffeomorphisms.
Int.J.Mod.Phys. {\bf A5} (1990) 833.}

\def\SI{Wilson, G.W.: The modified Lax and two-dimensional Toda Lattice
equations
associated with simple Lie Algebras. Ergod.Th. and Dynam.Sys. {\bf
1} (1981) 361}

\def\SM{Kac, V.G., Peterson, D.H.: 112 Constructions of the Basic
Representation
 of
the Loop Group of $E_{8}$. In: Symposium on Anomalies, Geometry and Topology.
Bardeen, W.A., White, A.R.(ed.s). Singapore: World Scientific 1985}

\def\RE{Kac, V.G., Wakimoto, M.: Exceptional Hierarchies of Soliton
Equations. Proceedings of Symposia in Pure Mathematics. Vol 49 (1989)
191}

\def\RG{De Groot, M.F., Hollowood, T.J., Miramontes, J.L.: Generalized
Drinfel'd-Sokolov Hierarchies. IAS and Princeton preprint IASSNS-HEP-91/19,
PUPT-1251 March 1991}

\def\RI{Gel'fand, I.M., Dikii, L.A.: Asymptotic Behaviour of the Resolvent
of Sturm-Louville Equations and the Algebra of the Korteweg-de Vries
Equation.
Russ. Math.Surv. {\bf30} (5) (1975) 77; Fractional Powers of Operators
and Hamiltonian Systems. Funkts.Anal.Pril. {\bf 10} (1976) 13;
Hamiltonian Operators and Algebraic Structures Connected with them.
Funkt.Anal.Pril. {\bf13} (1979) 13}

\def\RJ{Kupershmidt, B.A. Wilson G. : Modifying Lax Equations and the
Second Hamiltonian Structure. Invent. Math. 62 (1981) 403-436}

\def\RK{Babelon, O., Viallet, C.M.: Integrable Models,
Yang-Baxter Equation and Quantum Groups. Part 1. Preprint SISSA-54/89/EP May
1989}

\def\RL{Fateev, V.A., Zamolodchikov, A.B.: Conformal Quantum Field Theories
in Two-dimensions having $Z(3)$ Symmetry. Nucl.Phys. {\bf B280}[FS18] (1987)
644}

\def\RKirillov{Kirillov, A.A.: Elements of the Theory of Representations.
Springer-Verlag 1976}

\def\RShansky{Semenov-Tian-Shansky, M.: Dressing Transformations and
Poisson Group Actions. Publ.RIMS. {\bf 21} (1985) 1237}

\def\RM{Mathieu, P., Oevel, W.: The $W_3^{(2)}$ Conformal algebra and the
Boussinesq hierarchy, Laval University Preprint (1991)}

\def\NPrefmark#1{ [#1] }
\date={June, 1991}
\Pubnum{PUPT--1263, IASSNS-HEP-91/42}
\Ref\Rg{\RG} \Ref\Sc{\SC} \Ref\Ri{\RI} \Ref\Sg{\SG} \Ref\Si{\SI}
\Ref\Rj{\RJ} \Ref\Se{\SE} \Ref\Sb{\SB} \Ref\Sm{\SM}
\Ref\Rk{\RK} \Ref\Rkirillov{\RKirillov} \Ref\Rl{\RL}
\Ref\Rshansky{\RShansky} \Ref\Re{\RE} \Ref\Rm{\RM}

\endpage

\titlepage
\title{GENERALIZED DRINFEL'D--SOKOLOV HIERARCHIES II:\break
THE HAMILTONIAN STRUCTURES}
\author{Nigel J. Burroughs \& Mark F. de Groot}
\address{Institute For Advanced Study,\break
Olden Lane, Princeton, N.J. 08540.}
\andauthor{Timothy J. Hollowood \& J. Luis Miramontes}
\address{Joseph Henry Laboratories, Department of Physics,\break
Princeton University, Princeton, N.J. 08544.}
\abstract{
In this paper we examine the bi-Hamiltonian structure of the
generalized KdV-hierarchies.
We verify that both Hamiltonian structures take the form
of Kirillov brackets on the Kac-Moody algebra, and
that they define a coordinated system.
Classical extended conformal algebras are obtained from the
second Poisson bracket. In particular,
we  construct  the $W_n^{(l)}$ algebras, first
discussed for the case $n=3$ and $l=2$ by A. Polyakov and M. Bershadsky.}
\endpage

\chapter{Introduction}

This paper is a continuation of ref. [\Rg], where
we generalized the Drinfel'd-Sokolov construction
of integrable
hierarchies of partial differential equations
from Kac-Moody algebras, see [\Sc].
The work of Drinfel'd and Sokolov, itself, constituted
a generalization of the original
Korteweg-de Vries (KdV) hierarchy, the archetypal integrable
system. The main
omission from our previous paper was a discussion of the Hamiltonian formalism
of these integrable hierarchies, which is the subject of this present paper.
A Hamiltonian analysis of these integrable systems allows  a much deeper
insight into their structure, in particular
important algebraic structures are encountered such as
 the Gel'fand-Dikii algebras [\Ri], or
classical $W$-algebras, which arise as the second Hamiltonian
structure of the $A_n$-hierarchies of
Drinfel'd and Sokolov. Though we shall  say more about this later,
 the exemplar of this connexion is found in the original KdV hierarchy
whose second Hamiltonian structure is the
Virasoro algebra. The new hierarchies of [\Rg] lead amongst other
things to the
$W^{(l)}_N$-algebras for $1\le l\le N-1$, introduced in [\Sg].

A feature often encountered in the Hamiltonian analysis of integrable
hierarchies, is the presence
of two {\it coordinated\/} Poisson structures which we designate
$\{\phi,\psi\}_1$
and $\{\phi,\psi\}_2$. The property of {\it coordination\/} implies
that the one-parameter family of brackets
$$\{\phi,\psi\}=\{\phi,\psi\}_1+\mu\{\phi,\psi\}_2,$$
$\mu$ arbitrary, is also a Poisson structure, which is a non-trivial statement
as regards the Jacobi
identity. We say a system has a {\it bi-Hamiltonian
structure\/} if the brackets are coordinated and if the Hamiltonian flow can be
written in
two equivalent ways
$${\dot\phi}=\{H_2,\phi\}_1=\{H_1,\phi\}_2.$$
Under various general assumptions, the existence of a bi-Hamiltonian
structure implies the existence of an infinite hierarchy of flows,
that is, an infinite set of Hamiltonians $\{H_i\}$, such that
$$\partial_{t_i}\phi=\{H_{i+1},\phi\}_1=\{H_i,\phi\}_2,$$
where the Hamiltonians are in involution with respect to both
Poisson brackets, whence  the flows $\partial_{t_i}$ commute.
In the example of the KdV hierarchy, which has as
its first non-trivial flow the original KdV equation,

$${\partial u\over\partial {t_1}}=-{1\over4}u^{\prime\prime\prime}+
{3\over2}uu^\prime,$$
 where prime indicates differentiation with respect to $x$,
the two Poisson structures are
$$\eqalign{\{u(x),u(y)\}_1&=2\delta^\prime(x-y)\cr
\{u(x),u(y)\}_2&={1\over2}\delta^{\prime\prime\prime
}(x-y)-2u(x)\delta^\prime(x-y)-u^\prime(x)\delta(x-y).\cr}\eqn\kdv$$
One notices that the second structure is nothing but the Virasoro
algebra, as was already mentioned. There are  hierarchies that do not
admit a bi-Hamiltonian structure, for example the {\it modified\/} KdV
hierarchy (mKdV) (and its generalizations [\Sc,\Si])
which, as is well known, is related to the KdV
hierarchy by the {\it Miura Map\/} [\Sc,\Rj].
The Miura map takes a solution $\nu(x)$ of
the mKdV hierarchy  into a solution of the KdV hierarchy by
$$u(x)=-\nu^\prime(x)-\nu(x)^2.$$
This non-invertible mapping is in fact a Hamiltonian map from the single
Hamiltonian structure of the mKdV hierarchy to the
second Hamiltonian structure of the KdV hierarchy. The existence of a
{\it modified\/}
hierarchy associated to a KdV hierarchy is a feature also encountered in
the generalizations of [\Rg] and [\Sc].
In fact,
the situation is richer than this, since there exist a
tower of {\it partially modified\/} hierarchies (pmKdV),
at the top of which is the
KdV hierarchy and at the bottom its associated modified hierarchy
[\Rg]. Each of
these hierarchies has a Miura transform connecting it with the hierarchies
above. We shall show
that the KdV hierarchies of ref. [\Rg] admit a bi-Hamiltonian
structure, whereas for the partially modified hierarchies
we only obtain a  single Hamiltonian structure. The Miura map is proved to
be Hamiltonian, connecting the pmKdV hamiltonian structure to the second
Hamiltonian structure of the KdV hierarchy.

In order to make this paper reasonably self-contained,
we briefly review in section 2 relevant details of
ref. [\Rg], highlighting those
aspects which are important for the construction
of the Hamiltonian structures. Section 3 is the main body of the
paper, in which the two Poisson brackets are proposed and skew-symmetry and
the Jacobi identity are checked. In fact, the stronger statement of
coordination is proved. Section 4 discusses the way in which the
hierarchies lead to extended conformal algebras.
Section 5 discusses the partially modified KdV hierarchies and their
associated Miura mappings, proving in particular that the
Miura map is a Hamiltonian mapping.
Section 6 is devoted to applying the
preceding formalism to a number of examples. In particular, we
consider the Drinfel'd-Sokolov KdV hierarchies for the untwisted
Kac-Moody algebras, the fractional KdV hierarchy of ref. [\Se], and
various other cases.

\chapter{Review}

In this section we summarize certain salient aspects of
ref. [\Rg], to which one should refer for further details.

The central object in the construction of the hierarchies
is a Kac-Moody algebra $\hat
g$, realized as the loop algebra $\hat g=g\otimes{\bf
C}[z,z^{-1}]\oplus{\bf C}d$, where $g$ is a finite Lie algebra. The
derivation $d$ is chosen to induce the {\it homogeneous\/} gradation, so
that $[d,a\otimes z^n]=n\,a\otimes z^n$ $\forall\,a\in g$. One can
define other gradations as follows [\Sb]:

\noindent
{\bf Definition 2.1}. A {\it gradation of type ${\bf s}$\/}, is defined
via the derivation $d_{\bf s}$ which satisfies
$$[d_{\bf s},e_i\otimes z^n]=(nN+s_i)e_i\otimes z^n,$$
where $e_i$, $i=1,\ldots,{\rm rank}(g)$,
are the raising operators associated to the simple roots
of $g$, in some Cartan-Weyl basis of $g$, $N=\sum_{i=0}^{{\rm
rank}(g)}k_is_i$, where $k_i$ are the Kac Labels of $g$, and ${\bf
s}=(s_0,s_1,\ldots,s_{{\rm rank}(g)})$ is a vector of rank$(g)+1$ non-negative
integers.

Each derivation can be expressed in the following way
$$
d_s=N(d+\delta_{{\bf s}} \cdot H),\ \ \ \ \delta_{{\bf s}}
={1 \over N} \sum_{k=1}^{{\rm rank} (g)}
\left( {2 \over \alpha_k^2}\right) s_k \omega_k, \eqn\deriv
$$
where $\alpha_i$ are the simple roots of $g$, $H$ is the Cartan
subalgebra of $g$ and the $\omega_i$ are the
fundamental weights ($\alpha_i\cdot\omega_j=(\alpha_i^2/2)\delta_{ij}$).
Observe that the difference
$d_s-N d$ is an element of the Cartan subalgebra of $g$.

Under a gradation of type $\bf
s$, $\hat g$ is a ${\bf Z}$-graded algebra:
$$\hat g=\bigoplus_{i\in{\bf Z}}\hat g_i({\bf s}).$$
The homogeneous gradation corresponds to ${\bf s}_{\rm hom}\equiv(1,0,\ldots,
0)$.

An important r\^ole is played by the {\it Heisenberg subalgebras\/} of $\hat
g$,
which are maximal nilpotent subalgebras of $\hat g$, see ref. [\Sm] for a
definition. It is
known that, up to conjugation, these are in one-to-one correspondence
with the conjugacy classes of the Weyl group of $g$ [\Sm].
We denote
these subalgebras as ${\cal H}[w]$, where $[w]$ indicates the
conjugacy class of the Weyl group of $g$.

\noindent
{\bf Remark}. For an element $\Lambda\in{\cal H}[w]$, the Kac-Moody
algebra has the decomposition $\hat g={\rm Ker}({\rm
ad}\,\Lambda)\oplus{\rm Im}({\rm ad}\,\Lambda)$. In ref. [\Rg],
a distinction was made between hierarchies of type I and type II.
This referred to whether the element $\Lambda$ was {\it regular\/}, or
not ---
regularity implying that ${\rm Ker}({\rm ad}\,\Lambda)={\cal H}[w]$.
 In what follows we shall restrict ourselves to the former case.

\noindent
{\bf Remark}.
Associated to each Heisenberg subalgebra
there is a distinguished gradation of
type ${\bf s}$, which we denote ${\bf s}[w]$, with the property that ${\cal
H}[w]$ is an invariant subspace under ad$(d_{{\bf s}[w]})$ [\Rg].

One can introduce the notion of a partial ordering on the set
of gradations of type ${\bf s}$. We say ${\bf s}\succeq{\bf s}^\prime$ if
$s_i\neq0$ whenever $s^\prime_i\neq0$.

\noindent
{\bf Lemma 2.1}. [\Rg].
{\it An important property of this partial ordering is that
if ${\bf s}\succeq{\bf s}^\prime$ then the following is true

$(i)\ \  \hat g_0({\bf s})\subseteq\hat g_0({\bf s}^\prime)$,

$(ii)\ \ \hat g_j({\bf s})\subset\hat g_{\geq0}({\bf s}^\prime)\ {\rm or}\
\hat
g_{\leq0}({\bf s}^\prime)$, depending on whether $j>0$ or $j<0$
respectively,

$(iii)\ \ \hat g_j({\bf s}^\prime)\subset\hat g_{>0}({\bf s})\  {\rm or} \ \hat
g_{<0}({\bf s})$, depending on whether $j>0$ or $j<0$,
respectively.
}

In the above, we have used the notation $\hat g_{>a}({\bf s})=\oplus_{i>a}
\hat g_i({\bf s})$ and so on, to indicate subspaces of $\hat g$.

The construction of the hierarchies relies on the matrix Lax equation.
First of all, associated to the data $(\Lambda,{\bf s},[w])$
one defines the object
$$L=\partial_x+q+\Lambda,\eqn\Aa$$
where $\Lambda$ is a constant element of ${\cal H}[w]$ with well
defined positive ${\bf s}[w]$-grade $i$. By
constant we mean $\partial_x\Lambda=0$.
The fact that it is possible to choose $\Lambda$ to have a
well defined ${\bf s}[w]$-grade
follows from the second remark above.
The {\it potential} $q$ is
defined to be an element of $C^\infty({\bf R / Z},Q)$, where $Q$ is
the following subspace of $\hat g$:
$$Q=\hat g_{\geq0}({\bf s})\bigcap\hat g_{<i}({\bf s}[w]),\eqn\Ab$$
where ${\bf s}$ is any other gradation such that ${\bf s}\preceq{\bf s}[w]$.
The potentials are taken to be periodic functions, so as to avoid
technical complications [\Sc].

In this paper
our interest is principally in the KdV-type hierarchies, for which
the gradation ${\bf s}$ is the homogeneous gradation. For these systems the
analysis of Drinfel'd and Sokolov in ref. [\Sc] generalizes, leading to a
bi-Hamiltonian structure. Thus,
for brevity we introduce the following notation---superscripts
will denote ${\bf s}[w]$-grades, so that
$\hat g^j\equiv\hat g_j({\bf s}[w])$, and
subscripts will indicate homogeneous grade.

The function $q(x)$ plays the r\^ole of the phase space coordinate in
this system. However, there exist symmetries
in the system corresponding to the gauge transformation
$$L\rightarrow SLS^{-1},\eqn\Acc$$
with $S$ being generated by $x$ dependent functions on the
subalgebra $P\subset\hat g$, where
$$P=\hat g_0({\bf s})\bigcap\hat g_{<0}({\bf s}[w]).\eqn\Ad$$
The {\it phase
space\/} of the system ${\cal M}$ is the set of gauge equivalence
classes of operators of the form $L=\partial_x+q+\Lambda$.
The space of functions ${\cal F}$ on ${\cal M}$ is the
set of gauge invariant
functionals of $q$ of the form
$$\varphi[q]=\int_{{\bf R}/{\bf Z}}
dx\,f\left(x,q(x),q^\prime(x),\ldots,q^{(n)}(x),\ldots\right).$$
It is straightforward to find a basis for ${\cal F}$, the gauge
invariant functionals. One simply performs a non-singular
gauge transformation to take $q$ to some canonical form $q^{\rm can}$.
The components of $q^{\rm can}$ and their derivatives then provide the desired
basis. For
instance, for the generalized $A_n$-KdV hierarchies of Drinfel'd and
Sokolov, $q$ consists of
lower triangular $n+1$ by $n+1$ dimensional matrices, while the gauge
group is generated by strictly lower triangular matrices. A good gauge slice,
and the choice made in [\Sc], consists of matrices of
the form
$$\pmatrix{0&0&\cdots&0&0\cr \vdots&\vdots&&\vdots&\vdots\cr
0&0&\cdots&0&0\cr u_1&u_2&\cdots&u_n&0\cr}.\eqn\bas$$
The $u_i$'s and their
derivatives provide a basis for $\cal F$.

The outcome of applying the procedure of Drinfel'd and Sokolov to \Aa\
is
that there exists an infinite number of commuting flows on the gauge
equivalence classes of $L$. These flows have the following form. For
each element of the Centre of ${\rm Ker}({\rm ad}\,\Lambda)$ with
positive ${\bf s}[w]$-grade, which we denote by $K$,  there are
two gauge equivalent ways of writing the flows:
$${\partial L\over\partial t_b}=\left[A(b)^{\geq0},L\right],\ \ \ \ \ \
{\partial L\over\partial t^\prime_b}=\left[A(b)_{\geq0},L\right],$$
where the superscript and subscript $\geq0$ refer to projections onto
non-negative components in ${\bf s}[w]$-grade and $\bf s$-grade
respectively. In the case where $\Lambda$ is regular, $K={\cal
H}[w]^{>0}$ and so we can construct a flow for each element of the
Heisenberg algebra with positive grade (the type II
hierarchies require a somewhat different treatment).
The generator $A(b)$ is constructed from the Heisenberg algebra via
the transformation
$A(b)=\Phi^{-1}b\Phi$, where $b\in K$, and $\Phi=1+\sum_{j<0}\Phi^j$,
$\Phi^j\in C^\infty({\bf R / Z},{\rm Im}({\rm ad}\,\Lambda) \cap \hat g^j)$,
is the unique transformation which takes $L$ to
$${\cal L}=\Phi L\Phi^{-1}=\partial_x+\Lambda+\sum_{j<i}h^j,\eqn\Ae$$
where $h^j\in C^\infty({\bf R / Z},{\cal H}[w])$ with
${\bf s}[w]$-grade $j$.
The equations of motion take the following form in
the coordinates $q^{\rm can}$
$${\partial L^{\rm can}\over\partial t_b}=\left[A(b)_{\geq0}+\theta_b,L
^{\rm can}\right],$$
where $L^{\rm can}=L(q^{\rm can})$, and
$\theta_b\in C^\infty({\bf R}/{\bf Z},P)$
is the generator of an infinitesimal gauge transformation which
compensates for the fact that a flow will generically take $q$ out of
the gauge slice.

The quantities $h^j$ are the conserved densities for the flows, that
is, there exist quantities $a^j$ such that
$$\partial_th^j+\partial_x a^j=0.$$
These conserved densities are, in fact, the Hamiltonian densities for
the hierarchies.
In ref. [\Rg] it was shown that $a^j={\rm constant}$ for $j\geq0\ $
($j<i$), and
therefore the quantities $h^j$ for $i>j\geq0$ are constant under all flows
in the hierarchy. This is an important observation to which we return
in section 3.6.

\chapter{The Hamiltonian Structures}

In this section we explicitly construct the two coordinated
Hamiltonian structures of the KdV-type hierarchies (defined by the
requirement that $\bf s$ is the homogeneous
gradation).
The first hamiltonian structure is a direct generalization
of the first hamiltonian structure of the KdV hierarchy,
while the second
involves a classical $r$-matrix.
Our approach follows that of Drinfel'd and Sokolov [\Sc].

\section{Preliminaries}

For each $b\in{\cal H}[w]^{>0}$, there are four ways to write the flow:
$$\eqalignno{{\partial L\over\partial t_b}&=[A(b)^{\geq0},L]=-[A(b)^{<0},
L]&\eqnalign\Af\cr
{\partial L\over\partial t_b^\prime}&=[A(b)_{\geq0},L]=-[A(b)_{<0},L],
&\eqnalign\Ag\cr}$$
where, as before, $A(b)=\Phi^{-1}b\Phi$.
The flows defined by \Af\ and \Ag\ only differ by a gauge
transformation. Indeed, by applying lemma 2.1 we have
$$\eqalign{A(b)^{<0}&=A(b)_{<0}+A(b)_{0}^{<0}\cr
A(b)_{\geq0}&=A(b)^{\geq0}+A(b)_0^{<0},\cr}$$
and so the flows are related by the infinitesimal gauge transformation
generated by
$A(b)_0^{<0}\in C^\infty({\bf R / Z},\hat g_0\cap\hat
g^{<0})$:
$${\partial L\over\partial t_b}={\partial
L\over\partial {t_b^\prime}}-[A(b)_0^{<0},L].$$
The flows along $t_b$ and $t_b^\prime$ are, of course, identical on
the phase space ${\cal M}$.

There is a
natural inner product on the functions $C^\infty({\bf R}/{\bf Z},\hat g)$,
defined as follows
$$(A,B)=\int_{{\bf R}/{\bf Z}}\ dx\,\langle A(x),B(x)
\rangle_{\hat g},$$
where $\langle\ ,\ \rangle_{\hat g}$ is the Killing form of $\hat g$.
Explicitly
$$\langle a\otimes z^n,b\otimes z^m\rangle_{\hat g}=\langle
a,b\rangle_g\delta_{n+m,0},$$
where $\langle\ ,\ \rangle_g$ is the Killing form of $g$.
With respect to an arbitrary gradation we can express the inner
product in terms of the (suitably normalized) Killing form of the
finite Lie algebra $\hat g_0({\bf s})$:
$$(A,B)=\sum_{k\in{\bf Z}}\int dx\,\langle
A_k(x),B_{-k}(x)\rangle_{\hat g_0({\bf s})},$$
where $A_k$ and $B_k$ are the components of $A$ and $B$ of grade $k$
in the $\bf s$-gradation, and $\langle\ ,\ \rangle_{\hat g_0
({\bf s})}$ is the Killing form of the finite Lie algebra $\hat g_0({
\bf s})$. The inner product does not depend on the particular
gradation chosen, as long as the Killing forms of the finite algebras are
suitably normalized.

The first stage of the programme is to rewrite \Af\ and \Ag\ in
Hamiltonian form. In order to accomplish this, components of $A(b)$
have to be related to the Hamiltonians of the flows, which are in turn
constructed from the conserved densities $h^j$.

\noindent
{\bf Definition 3.1}. For a constant element $b \in {\cal H}[w]^{>0}$, we
define
the following functional of $q$:
$$H_b[q]=(b,h(q)),$$
where $h(q)=\sum_{j<i}h^j$ is the sum of the conserved densities of
\Ae.

Next we introduce the functional derivatives of  functionals
of $q$.

\noindent
{\bf Definition 3.2}. For a functional $\varphi$ of $q$ we define its
functional derivative $d_q\varphi\in C^\infty({\bf R}/{\bf
Z},\hat g_{\leq0})$ via
$$\left.{d\over d\varepsilon}\varphi[q+\varepsilon
r]\right\vert_{\varepsilon=0}
\equiv\left(d_q\varphi,r\right),$$
for all $r\in C^\infty({\bf R}/{\bf Z},Q)$.

Observe that the functional derivative $d_q\varphi$ is valued in
the subalgebra $\hat g_{\leq0}$. This is connected to the choice of the
space $Q=\hat g_{\geq0} \cap \hat g^{<i}$, and
is explained
by a group theoretic formulation of the generalized KdV hierarchy, which
will be discussed in another publication.

Since $r \in C^\infty({\bf R}/{\bf Z},Q)$, there is an ambiguity
in the definition of the functional derivative $d_q\varphi$
corresponding to the fact that terms in the annihilator of $Q$ are
not fixed by the definition. Thus $d_q\varphi$ is
defined up to terms in $\hat g^{\leq -i}$, the annihilator of $Q$ in
$\hat g_{\leq0}$.
In fact we can interpret the functional
derivative as taking values in the quotient algebra $\hat g_{\leq0} \big/
\hat g^{\leq -i}$. Part of the analysis of the Poisson structure
in later sections involves proving that the Poisson brackets are well
defined given that
$$d_q\varphi \in C^\infty\left({\bf R}/{\bf Z},{\hat g_{\leq0}
\big/ \hat g^{\leq -i}}\right),$$
{\it i.e.\/}
that terms in $\hat g^{\leq -i}$ do not contribute. The fact that the
second Poisson structure is well defined
is linked to gauge invariance.

The definition of the functional derivative, def. 3.2,
is related to the familiar notion of functional derivative in
the following way. If we introduce some basis
$\{e_\alpha\}$ for $\hat g$, with dual basis $\{e_\alpha^\star\} \in \hat g$
under the inner product,
then if $q=\sum q_\alpha e_\alpha$ the derivative
$$d_q\varphi=\sum_{\alpha}{\delta\varphi\over\delta
q_\alpha}e_\alpha^\star\ \ \ \ {\rm mod}\ C^\infty\left({\bf R}/{\bf Z},\hat
g^{\leq-i}\right),$$
where $\delta\varphi/\delta q_\alpha$ is the conventional definition
of a function derivative.

Now we present two central theorems.

\noindent
{\bf Theorem 3.1} {\it The
functional derivative of $H_b[q]$ is\/}:
$$d_{q}H_b=A(b)_{\leq0}\ \ \ \  {\rm mod}\ C^\infty\left({\bf R}/{\bf Z},
\hat g^{\leq -i}\right).$$

\noindent
{\it Proof}. Consider definition 3.2 of the functional derivative:
$$\eqalign{{d\over d\varepsilon}H_b[q+\varepsilon r]&={d\over d\varepsilon}
(b,h(q+\varepsilon r))\cr
&=\left(b,{d\over d\varepsilon}{\cal L}(\varepsilon)\right),\cr}\eqn\Aii$$
where ${\cal L}(\varepsilon)\equiv{\cal L}(q+\varepsilon r)$,
using
\Ae. Now we use the
relation ${\cal L}(\varepsilon)=\Phi(\varepsilon)L(\varepsilon)\Phi^{-1}
(\varepsilon)$ to evaluate
$${d\over d\varepsilon}{\cal L}(\varepsilon)=\Phi(\varepsilon)
r\Phi^{-1}(\varepsilon)
+\left[{d\Phi(\varepsilon)\over d\varepsilon}\Phi^{-1}(\varepsilon)
,{\cal L}(\varepsilon)\right].\eqn\Aiii$$
Substituting this into \Aii, and using the identity
$$(A,[B,C])=-(B,[A,C]),$$
along with the fact that $[{\cal L},b]=0$, we have
$$\left.{d\over d\varepsilon}H_b[q+\varepsilon
r]\right\vert_{\varepsilon=0}=
\left(b,\Phi r\Phi^{-1}\right)=\left(\Phi^{-1}b\Phi,r \right).$$
So finally
$$
d_{q}H_b[q]=\left(\Phi^{-1}b\Phi\right)_{\leq0}\ \ \ \
{\rm mod}\ C^\infty\left({\bf R}/{\bf Z},\hat g^{\leq-i}\right),
$$
as claimed.

\noindent
{\bf Lemma 3.1}.
{\it The quantities $A(b)$ occurring in the time evolution equations \Af,
\Ag\ possess the following symmetry\/}:
$$A(zb)_{k+1}^{j+N}=zA_k^j(b).$$

\noindent
{\it Proof}. $A(zb)=\Phi^{-1}(zb)\Phi=zA(b)$, and since $z$ carries
homogeneous grade 1 and ${\bf s}[w]$-grade $N$, the result follows
trivially.

\noindent
{\bf Remark.}
A special case is the relation $A(zb)_{\leq0}=zA_{<0}(b)$.
The fact that this relation only holds for the homogeneous gradation
is ultimately the reason why the KdV hierarchies, for which ${\bf
s}={\bf s}_{\rm hom}$, admit two Hamiltonian structures, whereas the
partially modified KdV hierarchies only exhibit a
single Hamiltonian structure.

\section{The First Hamiltonian Structure}

The First Hamiltonian Structure is derived by considering the equation
for the flow in the form
$${\partial L\over\partial t_b}=-\left[A(b)_{<0},L\right].$$
Recall that $q$ has ${\bf s}[w]$-grade in the range $-N+1$ to $i-1$.
Since the maximum ${\bf s}[w]$-grade of $L$ is $i$, it
is easy to see that the terms that are
needed to express the flow are only those components of $A(b)_{<0}$
with ${\bf s}[w]$-grade from $-N+1-i$ to $-1$.
{}From theorem 3.1 we have the relation $d_q H_{zb}=A(zb)_{\leq0}$,
and so using lemma 3.1 we may re-express these quantities in terms of
$A(b)_{<0}$:
$$z\sum_{k=-N+1-i}^{-1}A(b)_{<0}^k=d_q H_{zb}\ \ \ \ {\rm mod}\
C^\infty\left({\bf R}/{\bf Z},\hat g^{\leq-i}\right).$$
Therefore, the flow can be written as
$${\partial L\over\partial t_b}=-\left[{1\over z}d_qH_{zb}
,L\right]_{\geq0},\eqn\Ajj$$
where the restriction to homogeneous grade $\geq0$ is crucial, and
ensures that the right-hand side is contained in $C^\infty({\bf
R}/{\bf Z},Q)$, as required. Notice that the contribution from terms in the
functional derivative of grade less than $1-i$
cannot contribute to \Ajj\ because of the projection.

So for a functional $\varphi$ of $q$:
$${\partial\varphi\over{\partial t_b}}=\left(d_q\varphi,{{\partial q}\over
{\partial t_b}}\right)=-\left(d_q\varphi,z^{-1}\left[
d_qH_{zb},L\right]_{\geq0}\right).$$
Written in this form, the restriction to non-negative
homogeneous grade is redundant, being automatically ensured
because $d_q\varphi$ has strictly non-positive homogeneous
grade.

The candidate Poisson bracket for the First Hamiltonian structure is
thus
$$\{\varphi,\psi\}_1=-\left(d_q\varphi,z^{-1}\left[
d_q\psi,L\right]\right),\eqn\Aj$$
for two functionals of $q$. Of course, we must check that \Aj\ is a
well defined Poisson bracket, and we must also consider the r\^ole of
gauge invariance. This we shall do in sections 3.4 and 3.5.

\section{The Second Hamiltonian Structure}

The second Hamiltonian Structure results from considering the flow
written in
the form
$${\partial L\over\partial t_b}=\left[A(b)_{\geq0},L\right].$$
Firstly, we split $A(b)_{\geq0}$ into the terms of zero and positive
homogeneous grade:
$$A(b)_{\geq0}=A(b)_0+A(b)_{>0}.\eqn\Ak$$
The terms of zero grade, $A(b)_0$, can have ${\bf s}[w]$-grade between
$-N+1$ and $N-1$. The functional derivative $\left( d_{q}H_b\right)_0$
gives the component of $A(b)_0$ with ${\bf s}[w]$-grade between the
greater of $-i+1$ or $-N+1$, and $N-1$, {\it i.e.\/}
$$A(b)_0=\left( d_{q}H_b \right)_0+\Psi,$$
where $\Psi$ represents the sum of terms of homogeneous grade zero and
${\bf s}[w]$-grade less than $1-i$, which
is zero if $i\geq N$. The terms of positive homogeneous grade in
\Ak, can be re-expressed using lemma 3.1:
$$A(b)_{>0}=zA(z^{-1}b)_{\geq0}.$$
Collecting these results, we have
$${\partial L\over\partial t_b}=\left[\left( d_{q}H_b \right)_0,L
\right]+z\left[A(z^{-1}b)_{\geq0},L\right]+\left[\Psi,L
\right].\eqn\Al$$
We can ignore the term involving $\Psi$ since this is just a
gauge transformation.
Then we notice that the second term in \Al\ is equal to $z\partial
L/\partial t_{z^{-1}b}$, which we can express in terms of functional
derivatives of $H_b$ using the first Hamiltonian structure \Ajj. So
\Al\ becomes
$${\partial L\over\partial t_b}=\left[ \left( d_{q}H_b\right)_0,
L\right]-z\left[z^{-1}d_qH_b,L\right]_{\geq0}.\eqn\Amm$$
This can be written slightly differently by using $z[z^{-1}a,b]_{\geq0}=
[a,b]_{>0}$.

The candidate Poisson bracket on functionals of $q$ is thus
$$\{\varphi,\psi\}_2=\left(d_q\varphi,\left[
d_{q}\psi_0,L\right]-\left[d_q\psi,L\right]_{>0}\right).
\eqn\neanderthal$$
The above expression can be rewritten in a form that is more
suitable for our later discussions:
$$\{\varphi,\psi\}_2=\left(d_{q}\varphi_0,\left[d_{q}
\psi_0,L\right]\right)-\left(d_{q}\varphi_{<0},\left[d_{q}\psi_{<0},
L\right]\right),\eqn\Am$$
where we have used the fact that $d_q\varphi=d_{q}\varphi_0+d_{q}
\varphi_{<0}$, and that the inner product matches terms of opposite
grade. In the following sections we discuss the r\^ole of gauge
symmetry, and whether these brackets define a symplectic structure.

\section{Gauge Invariance}

It has already been mentioned in section 2 that the hierarchies
exhibit a gauge symmetry. More specifically the form of $L$ is
preserved under the transformation
$$L\mapsto SLS^{-1},\eqn\An$$
or equivalently
$$q\mapsto\tilde q= S(q+\Lambda)S^{-1}-\Lambda+S\partial_x
S^{-1},\eqn\Ao$$
where $S$ is an $x$-dependent element of the group generated by the
subalgebra $P=\hat g_0\cap\hat g^{<0}$.
For the KdV-type hierarchies considered in this section,
$P$ is of maximum dimension for a given conjugacy class $[w]$.
As discussed in [\Rg],
the flow equations of the hierarchy should be understood as equations
on the gauge equivalence classes of $L$ under \An.
For the generalized KdV hierarchy, we have proposed two Hamiltonian
structures. The fact that the hierarchies define dynamics on gauge
equivalence classes implies
that these Hamiltonian structures should respect the
gauge symmetry, {\it i.e.\/} that the
Poisson structure is well defined on gauge invariant functionals.
However, before proceeding with the discussion of gauge invariance,
it is necessary to prove that the brackets are actually
well defined as functionals of $q$, {\it i.e.\/} that the ambiguity
of the functional derivatives $d_q\varphi$ and $d_q\psi$, consisting
of terms in $C^\infty({\bf R}/{\bf Z},g^{\leq -i})$,
do not contribute to the bracket.
This is obtained as a corollary of the following lemma.

\noindent
{\bf Lemma 3.2}. {\it For $\Psi\in C^\infty({\bf R}/{\bf Z},\hat
g^{\leq-i})$:
$$\matrix{(i)&\ \ \ \ \ \left(\Psi,z^{-1}\left[d_q\varphi
,L\right]\right)=0,\cr
(ii)&\ \ \ \ \ \left(\Psi_{<0},\left[d_{q}\varphi_{<0},L\right]\right)=
0,\cr
(iii)&\ \ \ \ \ \left(\Psi_0,\left[d_{q}\varphi_0
,L\right]\right)=0.\cr}$$
}

\noindent
{\it Proof}. The maximum ${\bf s}[w]$-grade of $L$ is $i$, and that of
$d_q\varphi$ is $N-1$, therefore the maximum ${\bf
s}[w]$-grade of $z^{-1}[d_q\varphi,L]$ is $i-1$. Since
$\Psi$ only has ${\bf s}[w]$-grade less than or equal to $-i$, the first part
of the lemma follows. To prove the second part of the lemma we notice
that the maximum ${\bf s}[w]$-grade of $d_{q}\varphi_{<0}$
is $-1$, by lemma 2.1. Therefore the maximum grade of the second
term in the right side of the inner product is $i-1$, showing that the
expression vanishes. To show that
the third expression vanishes is
more subtle and relies on the fact that the
brackets are properly defined on gauge invariant functionals, a point
that we shall come to shortly.
Notice, first of all, that the projection $\Psi_0$ is the generator of
an infinitesimal gauge transformation. The variation of $q$ under this
transformation is $\delta_\varepsilon q=\varepsilon[\Psi_0,L]$.
Since $\varphi[q]$ is gauge invariant we have
$$0=\left.{d\over d\varepsilon}\varphi[q+\delta_\varepsilon
q]\right\vert_{\varepsilon=0}=\left(d_q\varphi,[
\Psi_0,L]\right).\eqn\Ap$$
But $\Psi_0$ has ${\bf s}[w]$-grade $\leq-i$, therefore $[\Psi_0,L]$ has
${\bf s}[w]$-grade less than or equal to zero, so the only term which can
contribute to the right-hand side of \Ap\ is
$$\left(d_{q}\varphi_0
,[\Psi_0,L]\right)=-\left(\Psi_0,\left[d_{q}\varphi_0,L
\right]\right),$$
which follows from the invariance of the Killing form.
But this is zero by \Ap, and so the lemma is proved.

We now establish how the functional derivatives transform under gauge
transformations.

\noindent
{\bf Lemma 3.3}. {\it Under the gauge transformation \Ao, the
functional derivative of a gauge invariant functional $\varphi$
transforms as\/}:
$$d_q\varphi\mapsto Sd_q\varphi
S^{-1}.$$

\noindent
{\it Proof}. Consider the definition of the functional derivative
$$\left.{d\over d\varepsilon}\varphi[q+\varepsilon r]\right\vert_{
\varepsilon=0}=\left(d_q\varphi,r\right),$$
for constant $r\in Q$.
Since $\varphi[q]$ is a gauge invariant functional,
we perform the gauge transformation
$\tilde q+\varepsilon r\mapsto q+\varepsilon S^{-1}rS$
and obtain
$$\eqalign{\left.{d\over d\varepsilon}\varphi[\tilde q+\varepsilon
r]\right\vert_{\varepsilon=0}\equiv
\left.{d\over d\varepsilon}\varphi[q+\varepsilon
S^{-1}rS]\right\vert_{\varepsilon=0}&=\left(d_q\varphi
,S^{-1}rS\right)\cr &=\left(Sd_q\varphi
S^{-1},r\right),\cr}$$
using the ad-invariance of the inner product. Therefore, from the
definition of the functional derivative we have
$$d_{\tilde q}\varphi=Sd_q\varphi
S^{-1}.$$
Remember that the functional derivatives are only defined modulo terms of
${\bf\rm s}[w]$-grade less than $1-i$, and it is in this sense that the
equality
holds.

\noindent
{\bf Proposition 3.1}. {\it The Poisson brackets \Aj\ and \Am\ of two gauge
invariant functionals of $q$ are gauge invariant functionals of $q$.}

\noindent
{\it Proof}. For the first Poisson bracket, \Aj, the transformed bracket is
$$\left(d_{\tilde q}\varphi,z^{-1}\left[d_{\tilde q}
\psi,\tilde L\right]\right)=
\left(Sd_q\varphi S^{-1},z^{-1}\left[Sd_q
\psi S^{-1},SLS^{-1}\right]\right)=
\left(d_q\varphi,z^{-1}\left[d_q
\psi,L\right]\right),$$
where the last manipulation follows from the ad-invariance of the inner
product. The proof for the second Poisson
bracket proceeds in the same spirit, although in this case it also depends on
the fact that $S$ has zero homogeneous grade.

\section{The Jacobi Identity}

In this section we verify that both \Aj\ and \Am\ define Poisson brackets.
This entails checking that the brackets are skew symmetric
and that the Jacobi identity is satisfied. In fact,
we shall prove the stronger statement that they are coordinated.

In order to demonstrate that the Jacobi identity is satisfied,
we first make the following digression.
Consider a Lie algebra $g$, with a Lie
bracket denoted $[\ ,\ ]$.
Suppose we have an endomorphism $R\in{\rm End}\,g$, then we can
define a new bracket operation
$$[x,y]_R=[Rx,y]+[x,Ry],\eqn\Ar$$
$\forall$ $x$,$y\in g$, see [\Rk]. If the Jacobi identity is satisfied in $[\
,\
    ]_R$
then there exists a new Lie algebra structure on the underlying vector
space of $g$, denoted $g_R$.
The Jacobi identity translates into the
following condition on $R$
$$[Rx,Ry]-R([Rx,y]+[x,Ry])=\lambda[x,y],\eqn\As$$
for some proportionality constant $\lambda$.
This equation is known as the {\it modified
Yang-Baxter Equation\/} (mYBE) [\Rk].

The simplest example of this procedure is
when $g$ has the vector space decomposition
$g=a+b$, where $a$ and $b$ are subalgebras of $g$. If $P_a$ and $P_b$ are the
projectors onto these subalgebras, then we can define the new Lie algebra
$g_R$ via $R=(P_a-P_b)/2$. In this case
$$[x,y]_R=[P_ax,P_ay]-[P_bx,P_by],$$
which implies $g_R\simeq a\oplus b$, the mYBE then being satisfied with
$\lambda=-{1\over 4}$.

Applying this formalism to our situation, we consider the
vector-space decomposition $\hat g_{\leq0}=\hat g_0+\hat g_{<0}$, into
the subalgebras $\hat g_0$ and $\hat g_{<0}$ of the Lie algebra $\hat
g_{\leq0}$.
The importance of this decomposition is that the second
Hamiltonian structure may be succinctly rewritten as
$$\{\varphi,\psi\}_2=\left(q+\Lambda,\left[d_q\varphi,d_q\psi\right]_R\right)
-\left(d_q\varphi,\left(d_q\psi\right)^\prime\right),$$
where $R=(P_0-P_{<0})/2$, half the difference of the projector onto the
subspace of zero homogeneous grade and the projector onto the subspace
of strictly negative homogeneous grade. Notice that only the terms of
zero homogeneous grade,
$d_{q}\varphi_0$ and $d_{q}\psi_0$ contribute to the last term.

In fact, we may combine the first and second Poisson brackets into one
elegant expression using the following lemma.

\noindent
{\bf Lemma 3.4}. {\it The one-parameter family of endomorphisms of the
Lie algebra $\hat g_{\leq0}$ defined by
$$R_\mu=R-\mu\cdot{1\over z},$$
where $R=(P_0-P_{<0})/2$ and $\mu\in{\bf C}$,
satisfies the modified Yang-Baxter Equation.}

\noindent
{\it Proof}. It is useful to define $\sigma=-\mu/z$, with the property
that $\sigma:\ \hat g_{\leq0}\rightarrow\hat g_{<0}$. In order to
prove the lemma we must demonstrate that $R_\mu$ satisfies the mYBE.
The left-hand side of \As\ is equal to
$$\eqalign{[(R+\sigma)x,(R+\sigma)y]-(R+\sigma)\left([(R+\sigma)x,y]+
[x,(R+\sigma)y]\right)\cr
=[Rx,Ry]-R([Rx,y]+[x,Ry])-\sigma^2([x,y])-2R\sigma([x,y]),\cr}$$
where we used the fact that $[\sigma(x),y]=\sigma([x,y])$. Now, $-2R$
acts as the identity on $\sigma([x,y])$, and hence the above
expression is equal to
$$-\left({1\over 2}-{\mu\over z}\right)^2[x,y],$$
verifying that the mYBE is satisfied by $R_\mu$.

Since the endomorphism $R_\mu$  satisfies the mYBE,
there exists a Poisson bracket on the dual space given by the Kirillov
bracket construction [\Rkirillov].
Up to a term involving a derivative, which can be
interpreted as a central extension of $\hat g_0$
and causes no problem in the proof of the Jacobi identity,
this is the previously constructed
bi-Hamiltonian structure on the space of gauge invariant functions on
$Q$, equations \Aj, \Am. We
summarize the Hamiltonian structure in the form of a theorem.

\noindent
{\bf Theorem 3.2}.
{\it There is a one parameter family of Hamiltonian structures on
the gauge equivalence classes of the generalized KdV hierarchy given
by\/}
$$
\{\varphi,\psi\}_\mu=\left(q+\Lambda,\left[d_q\varphi,d_q\psi\right]_{R_\mu}
\right)-\left(d_q\varphi,\left(d_q\psi\right)^\prime\right),\eqn\Au
$$
{\it where $[\ ,\ ]_{R_\mu}$ is the Lie algebra commutator
constructed from $R_\mu=(P_0-P_{<0})/2-\mu/z$.
Expanding in powers of $\mu$, $\{\ ,\ \}_\mu=\mu\{\ ,\ \}_1+\{\ ,\ \}_2$,
we obtain the
two coordinated Hamiltonian structures on ${\cal M}$\/}
$$\eqalign{
\{\varphi,\psi\}_1=&-\left(d_q\varphi,z^{-1}\left[
d_q\psi,L\right]\right),\cr
\{\varphi,\psi\}_2=&\left(q+\Lambda,\left[d_q\varphi,d_q\psi\right]_{R}
\right)-\left(d_q\varphi,\left(d_q\psi\right)^\prime\right),\cr}
$$
{\it where $R=(P_0-P_{<0})/2$.
Under time evolution in the coordinate $t_b$,
the following recursion relation holds\/}:

$${\partial\varphi\over\partial t_b}=
\{\varphi,H_{zb}\}_1=\{\varphi,H_b\}_2.\eqn\Av$$

Recall that our analysis has concentrated on the
KdV-type hierarchies defined by the two
gradations $({\bf s}_{{\rm hom}}, {\bf s}[w])$.
In obtaining the
Poisson brackets from the dynamical equations, sections 3.1 and 3.2,
we have employed
special properties of the homogeneous gradation.
This dependence on
the homogeneous gradation can be observed in the formulae for the
Poisson brackets, the first Poisson structure involving a
factor of $z^{-1}$ while the second is expressed in terms of
the $R$-operator $R=(P_0-P_{<0})/2$. However,
gauge invariance removes explicit dependence of
the second Poisson bracket on the homogeneous gradation.
More explicitly, it is possible to express the second Poisson
bracket in terms of an arbitrary gradation ${\bf s}$, satisfying the
inequalities
${\bf s}_{{\rm hom}} \preceq {\bf s}
\preceq {\bf s}[w]$. This is accomplished through the
use of the following lemma

\noindent
{\bf Lemma 3.5.}
{\it Consider a Lie algebra $g$ with the subalgebras $A,C,A+B,B+C$.
Then if $R_A=(P_A-P_{B+C})/2,\ R_C=(P_{A+B}-P_C)/2$, the Lie algebra
commutators
satisfy\/}:
$$
[X,Y]_{R_C}=[X,Y]_{R_A}+[P_BX,Y]+[X,P_BY],\ \forall X,Y \in g.
$$

The proof of this lemma is just a question of writing out the Lie
brackets, and so is omitted. In actual
fact, we are interested in a centrally extended version of
this lemma applied to the centrally extended Lie algebras $\left( [\ ,\
]_{R},\omega_{\rm hom}
\right),\ \left( [\ ,\ ] _{R[{\bf s}]}, \omega_{\bf s}\right)$, where
$\omega_{\bf s}$ is the
central extension of $\hat g_{0}({\bf s})$,
$\omega_{\bf s}(X,Y)=\left(X^\prime,P_{0[{\bf s}]}Y \right)$.
With this modification, the lemma still holds, the additional terms
involving the central extension
$\omega(X,Y)=\left(X^\prime,Y\right)$.
We have the following proposition.

\noindent
{\bf Proposition 3.2}.
{\it The second Poisson bracket between gauge invariant functions
can be expressed in the form}
$$
\{\varphi,\psi\}_2=\left(q+\Lambda,\left[d_q\varphi,d_q\psi\right]_{R[{\bf
s}]}
\right)-\left(P_{0[{\bf s}]}d_q\varphi,\left(d_q\psi\right)^\prime\right),
$$
{\it where $R[{\bf s}]=(P_{\geq0[{\bf s}]}-P_{<0[{\bf s}]})/2$,
with the arbitrary gradation ${\bf s}$ satisfying
${\bf s}_{{\rm hom}} \preceq {\bf s}
\preceq {\bf s}[w]$.}

\noindent
{\it Proof}. The proof follows from lemma 3.5 with
$A=\hat g_0 \cap \hat g_{\geq0}({\bf s}),\
B=\hat g_0 \cap\hat g_{<0}({\bf s}) \subset \hat g_0 \cap\hat g^{<0},
C=\hat g_{<0}$, along with the fact that
the additional terms vanish owing to the gauge
invariance of functionals.

The importance of this
Proposition will become apparent in our later analysis of the
partially modified KdV hierarchies, [\Rg], in section 5,
for which the gradation ${\bf s}$ is chosen to be more general than
the homogeneous gradation hitherto considered.

\section{Centres}

In this section we point out that the Poisson brackets defined
in theorem 3.2
sometimes admit non-trivial centres. The existence of these
centres is directly related to the
Hamiltonian densities $h^j$ with $i>j\geq0$. As we have
already remarked these densities
are constant under all the flows of the hierarchy, and so not all the
functionals on ${\cal M}$ are dynamical.
We shall show below that the densities
are centres of the Poisson bracket algebra.

Before we proceed to the proposition we first establish a
useful lemma.

\noindent
{\bf Lemma 3.6}. {\it The functional $\Theta_f=(f,h(q))$, where $h(q)$
was defined in definition 3.1, for
$$f\in C^\infty({\bf R}/{\bf Z},\oplus_{j=1-i}^0{\cal H}^j[w]),$$
satisfies
$$d_q\Theta_f=\Phi^{-1}f\Phi\ \ \ \ {\rm mod}\ C^\infty\left(
{\bf R}/{\bf Z},\hat g^{\leq-i}\right).$$
}

\noindent
{\it Proof}. One follows the steps of theorem 3.1 up to equation \Aiii,
$$\left.{d\over d\varepsilon}\Theta_f[q+\varepsilon
r]\right\vert_{\varepsilon=0}=\left.\left(f,\Phi
(\varepsilon)r\Phi^{-1}(\varepsilon)+\left[{d\Phi(
\varepsilon)\over d\varepsilon}\Phi^{-1}(\varepsilon),{\cal L}(\varepsilon
)\right]\right)\right\vert_{\varepsilon=0}.\eqn\don$$
However, in this case $f$ is not a constant and so \don\ equals
$$\left(\Phi^{-1}f\Phi,r\right)+\left.\left(f^\prime,{d\Phi(\varepsilon)
\over
d\varepsilon}\Phi^{-1}(\varepsilon)\right)\right\vert_{\varepsilon=0
}.$$
The second term cannot contribute because
$(d\Phi(\varepsilon)/d\varepsilon)\Phi^{-1}(\varepsilon)$ has ${\bf
s}[w]$-grade $<0$ and $f^\prime$ has ${\bf s}[w]$-grade $\leq0$, and
so the lemma is proved.

\noindent
{\bf Proposition 3.3}. {\it The functionals of the form\/}
$\Theta_f=(f,h)$ {\it for\/}
$$f\in C^\infty({\bf R}/{\bf Z},\oplus_{j=1-i}^k{\cal
H}^j[w]),$$
{\it are centres of the first Poisson bracket algebra, for
$k=0$, and centres of the second Poisson bracket algebra, for
$k=-1$.}

\noindent
{\it Proof}. Lemma 3.6 implies that $d_q\Theta_f=\Phi^{-1}f\Phi$,
modulo terms of ${\bf s}[w]$-grade less than $1-i$, which will not
contribute to the Poisson brackets owing to lemma 3.2. We have
$$\{\varphi,\Theta_f\}_1=-\left(d_q\varphi,z^{-1}\left[
\Phi^{-1}f\Phi,L\right]\right).$$
This is zero because
$[\Phi^{-1}f\Phi,L]=-\Phi^{-1}f^\prime\Phi$, using the
definition of $\Phi$ in
section 2, equation \Ae,
and the fact that $z^{-1}\Phi^{-1}f^\prime\Phi$ has homogeneous
grade $<0$. This proves that $\Theta_f$ is a centre of the
first Hamiltonian structure. For the second structure
$$\eqalign{\{\varphi,\Theta_f\}_2&=\left(d_{q}\varphi_0,\left[\left(
\Phi^{-1}f\Phi
\right)_0,L\right]\right)-\left(d_{q}\varphi_{<0},\left[\left(\Phi^{-1}f\Phi
\right)_{<0},L\right]\right)\cr
&=\left(d_q\varphi,\left[\left(\Phi^{-1}f\Phi\right)_0,L\right]\right)+
\left(d_{q}\varphi_{<0},\Phi^{-1}f^\prime\Phi\right),\cr}\eqn\Axx$$
which follows because
$\left(\Phi^{-1}f\Phi\right)_{<0}=\Phi^{-1}f\Phi-\left(\Phi^{-1}f\Phi\right)_0$.
The second term in \Axx\ is zero owing to mismatched homogeneous grade.
If the ${\bf s}[w]$-grade of $f$ is less than zero,
the first term above induces a gauge transformation of
$\varphi$, which is zero because $\varphi$ is a gauge invariant
functional, and so the proposition is proved. Notice that the
proof for the second structure does not cover the case when $f$ has
zero ${\bf s}[w]$-grade.

It is an obvious corollary of the proposition that the densities $h^j$, for
$0\leq j<i$ are non-dynamical, as was proved in ref. [\Rg] directly.
It is interesting to notice that $h^0$ is {\it not\/} a
centre of the second Poisson bracket algebra, even though it is
non-dynamical, a point which will be apparent in the examples
considered in section 6.

\chapter{Conformal Symmetry}

It is proved in
[\Sc] that the KdV hierarchies
exhibit a scale invariance, {\it i.e.\/} under the transformation
$x\mapsto\lambda x$, for constant $\lambda$, each quantity in the equations
can be assigned a scaling dimension such that the equations are
invariant. The original KdV equation provides
a typical example; the appropriate transformations are $x\mapsto\lambda
x$, $u\mapsto\lambda^{-2}u$ and $t\mapsto\lambda^3t$.
In this section, we prove that this scaling invariance
generalizes to the hierarchies defined in [\Rg], and
are, in fact, symmetries of the second symplectic structure.
By generalizing this  result to arbitrary conformal (analytic)
transformations, $x\mapsto y(x)$, we surmise that
the second Poisson bracket algebra contains (as a subalgebra) the
algebra of conformal transformations, {\it i.e.\/} a Virasoro algebra.
This would imply that the second
Poisson bracket algebra is  an extended chiral conformal algebra,
generalizing the occurrence of the $W$-algebras as the second Poisson
bracket algebra of the hierarchies of Drinfel'd and Sokolov.

Consider the transformation $x \rightarrow \lambda x$ on the Lax operator
$L=\partial_x +\Lambda+q(x)$. In order that this rescaling can be lifted to
a symmetry of the equations of motion, it is necessary that the
form of $L$ is preserved. To this end we consider the transformation
$z \rightarrow \tilde z =\lambda^{-{N / i}} z$,
with a simultaneous adjoint action by:
$$
U=\lambda^{-(N/i)\delta_{{\bf s}[w]} \cdot H}
\equiv\exp\left( {-{N \over i}\log \lambda \ \delta_{{\bf s}[w]} \cdot
H}\right),
\eqn\Umatrix
$$
where $\delta_{{\bf s}[w]}$ is defined in \deriv. By means of this
transformation we are lead to the following proposition.

\noindent
{\bf Proposition 4.1}
{\it The dynamical equation of the hierarchy
generated by the Hamiltonian
$H_{b^j}$ (with respect to the second Hamiltonian structure), is invariant
under the transformations
$$x\mapsto\lambda x, \
t_{b^j}\mapsto\lambda^{j/i}t_{b^j}, \
q^k\mapsto\lambda^{k/i-1}q^k,
$$
where $i$ is the ${\bf s}[w]$-grade of $\Lambda$, and $q^k$ is the
component of $q$ with ${\bf s}[w]$-grade $k$.
}

\noindent
{\it Proof}. We consider the following transformation of $L$
$$L(x,q;z)=\lambda UL(y,\tilde q;z\lambda^{-N/i})U^{-1},$$
where $y=\lambda x$. First of all notice that under the transformation
$$\Lambda(z)=\lambda U\Lambda(\tilde z)U^{-1},$$
which ensures that $L$ preserves its form. Secondly, under this
transformation the coefficient  of $q$ with ${\bf
s}[w]$-grade $k$ transforms as $\tilde q^k=\lambda^{k/i-1}q^k$.
In order to derive the rescaling of the time parameter $t_{b^j}$,
consider the change in the transformation $L=\Phi^{-1} {\cal L} \Phi$ in
equation \Ae.
If $\tilde\Phi$ is the transformation which conjugates $L(y,\tilde q;\tilde
z)$ into the Heisenberg subalgebra, then the relationship between
$\Phi$ and $\tilde\Phi$ is
$$\Phi(x,q;z)=U\tilde\Phi(y,\tilde q;\tilde z)U^{-1},$$
because adjoint action by $U$ preserves the decomposition $\hat g=
{\rm Im}({\rm ad} \Lambda) \oplus {\rm Ker}({\rm ad} \Lambda)$,
and the eigenspaces $\hat g^j$.
Therefore $\left(\Phi^{-1}b^j\Phi\right)=\lambda^{j/i}U\left(\tilde
\Phi^{-1}\tilde b^j\tilde\Phi\right)U^{-1}$, where $\tilde b^j$ is equal to
$b^j$ with $z$ replaced by $\tilde z$. Since adjoint action by $U$
commutes with the projections $P_{\geq0},\ P_{<0}$,
we deduce that the time evolution equations, \Af, \Ag, are
invariant if $\tilde t_{b^j}=\lambda^{j/i}t_{b^j}$.

Notice that the cumulative effect of the transformation on $z$ and the
conjugation by $U$ is equivalent to a global rescaling of the gradation
${\bf s}[w]$, {\it i.e.\/} action by $\exp(-\log\lambda\,{\rm ad}(d_{
{\bf s}[w]})/i)$.

We have shown that the equations of the hierarchy are quasi-homogeneous
under the lift of the transformation $x\mapsto\lambda x$. However one can
lift the more general conformal transformations $x\mapsto y(x)$ onto the
phase space in a similar manner, by altering the global ${\bf
s}[w]$-rescaling, to a local rescaling. Thus we
transform $z \mapsto \tilde z=
(y')^{-N/i}z$, and alter \Umatrix\ to:
$$
U[y]=(y')^{-(N/i)\delta_{{\bf s}[w]} \cdot H},
$$
where $y'=dy/dx$. The transformation of the Lax operator is now
$$
L(x,q;z)=y'U[y]L(y,\tilde q;\tilde z)U[y]^{-1},
$$
corresponding to the following transformation of the potential $q$
$$
q(x;z)=y'U[y]\tilde q(y;\tilde z)U[y]^{-1}+
{N \over i} \left({y''\over y'}\right){\delta_{{\bf s}[w]} \cdot H}.
\eqn\trans$$
This conformal transformation induces a corresponding
conformal transformation on the space of gauge equivalence classes,
{\it i.e.\/} on ${\cal M}$. This follows because
the gauge group is generated by $\hat g_0 \cap \hat g^{<0}$, and under a
${\bf s}[w]$-rescaling this subalgebra is invariant. Thus if
$L = S^{-1}L^\prime S$, with $S$ in the gauge group, then under a
conformal transformation the corresponding Lax operators are
related by the gauge transformation defined by $U[y]^{-1}SU[y]$.
We observe that under a conformal transformation, the non-dynamical
functionals on ${\cal M}$ are also subject to transformation.

With the conformal transformation $x \rightarrow y(x)$, the
manipulations in the proof of proposition 4.1 can now be reproduced,
with $\lambda$ replaced by $y'$,
however, one now finds the flow variables transform in a more
complicated way:
$$t_{b^j}\mapsto\tilde t_{b^j}=\left( y'\right)^{j/i}t_{b^j}+\cdots,$$
where the dots represent terms which depend on the variables
$t_{b^k}$, with $k<j$, which vanish for the scale transformation.

\noindent
{\bf Example.} Let us consider the case of the original KdV
hierarchy. We want to determine how the gauge invariant function $u$,
defined in section 2, transforms under the transformation \trans. The form
of $L$ in this case is
$$L=\partial_x+\pmatrix{a&0\cr b&-a\cr}+\pmatrix{0&1\cr z&0\cr},$$
and the gauge invariant function is $u=a^2+b-a'$. Under the transformation
\trans\ one finds
$$a=y'\tilde a+{y''\over2y'},\ \ \ b=(y')^2\tilde b,$$
and so
$$\eqalign{u&=(y')^2(\tilde a^2+\tilde b-\partial_y\tilde a)-{y'''\over2
y'}+{3\over4}\left({y''\over y'}\right)^2\cr
&=(y')^2\tilde u-{1\over2}{\cal S}(y),\cr}$$
where ${\cal S}(y) = y'''/y'-{3\over2}(y''/y')^2$ is the Schwartzian
derivative of $y$ with respect to $x$. Thus $u(x)$ transforms like a
projective connexion or Virasoro generator. This gives a hint of a hidden
conformal symmetry in the system.

Below, we show that the second
Poisson bracket algebra, for a general hierarchy,
is invariant under an arbitrary conformal transformation.

We now wish to determine how the functional derivatives of the gauge
invariant functionals transform under the conformal transformation \trans.
Recall that the functional derivative $d_q \varphi$
of a functional $\varphi \in {\cal F}$ is a gauge invariant
element of $C^\infty({\bf R}/{\bf Z}, \hat g_{\leq0} \big/ g^{\leq -i})$.
We use the notation $d_q \varphi(x;z)$ for the
functional derivative, explicitly indicating the integration variable,
$x \in {\bf R} / {\bf Z}$ and loop variable $z$ in
definition 3.2, {\it i.e. }
$$
\left.{d\over d\varepsilon}\varphi[q+\varepsilon r]\right\vert
=\int dx\langle d_{q}\varphi(x;z),r(x;z)\rangle,\eqn\mad
$$
for arbitrary $r\in C^\infty({\bf R}/{\bf Z},Q)$.
Our notation mirrors that of the finite dimensional case:
the functional derivative $d_q \varphi$ taking values
in the `cotangent space' at $q \in {\cal M}$.
In the calculation of the functional derivative $d_q \tilde \varphi$,
it is necessary to take proper account of the variables $x$ and
$z$ in the functional derivative; the
relationship between $d_q \tilde \varphi$ and $d_{\tilde q} \varphi$
involving
a transformation of these variables similar to that occurring in \trans.

\noindent
{\bf Lemma 4.1} {\it The transformation \trans\ induces
the following transformation on the functional
derivatives
$$
d_{q}\tilde\varphi (x;z)
=U[y] d_{\tilde q}\varphi (y(x);\tilde z) U[y]^{-1},
$$
where $\tilde\varphi[q]=\varphi[\tilde q]$ is the pull-back of
the gauge invariant functional $\varphi$.}

\noindent
{\it Proof\/.} Consider the definition of $d_{q}\varphi(x;z)$ in \mad.
Making the dependence on all the variables explicit,
under \trans\ we have
$\tilde \varphi[q+\varepsilon r]=\varphi
[\tilde
q+\varepsilon \tilde r]$,
with $\tilde r(y(x); \tilde z)=(y')^{-1}U[y]^{-1}r(x;z)U[y]$,
and so from \mad\ we
deduce
$$\eqalign{\int dx\langle d_{q}\tilde \varphi(x;z),r(x;z)\rangle&=\int
 dy(y')^{-1}\langle
d_{\tilde q} \varphi (y;\tilde z),U^{-1}r(x;z)U\rangle\cr
&=\int dx\langle Ud_{\tilde q}\varphi (y;\tilde z) U^{-1},r(x;z)\rangle,\cr}$$
hence the lemma is proved.

\noindent
{\bf Proposition 4.2}. {\it The transformation \trans\ is a Poisson
mapping of the second Poisson structure.}

\noindent
{\it Proof}. We have to show that $\{\tilde\varphi,\tilde\psi\}_2
[q]=\{\varphi,\psi\}_2[\tilde q]$. First of all, $U[y]$ carries zero
grade, therefore from lemma 4.1 $d_{q}\tilde\varphi (x;z)_0=Ud_{\tilde
q} \varphi (y;\tilde z)_0 U^{-1}$;
similarly for $d_{q}\tilde\varphi (x;z)_{<0}$.
This means, using the expression \neanderthal\ and lemma 4.1
$$
\eqalign{\{\tilde\varphi,\tilde\psi\}_2[q]&=\int dx\left\langle Ud_{\tilde q}
\varphi (y;\tilde z)_0U^{-1},\left[Ud_{\tilde q}\psi (y;\tilde z)_0U^{-1},y'U
L(\tilde q)U^{-1}\right]\right\rangle\cr
&\ \ \ \ \ \ \ -\int dx\left\langle Ud_{\tilde q}
\varphi (y;\tilde z)_{<0}U^{-1},\left[Ud_{\tilde q}
\psi (y;\tilde z)_{<0}U^{-1},y'UL(y,\tilde q;\tilde z)U^{-1}
\right]\right\rangle\cr
&=\{\varphi,\psi\}_2[\tilde q],}$$
owing to the ad-invariance of the Killing form, and the fact that the
factor of $y'$ transforms the measure in just the right
way, $dx\mapsto dy$.
It is important that the inner product on the
Kac-Moody algebra pairs terms of opposite grade so that it is invariant
under the transformation $z\mapsto\tilde z$, {\it i.e.\/}
$$\langle A(z),B(z)\rangle=\langle A(\tilde z),B(\tilde z)\rangle.$$

The fact the the first symplectic structure
does not respect the conformal symmetry is due
to the presence of the $z^{-1}$ term in (3.6).

Since the second Poisson bracket is preserved by the conformal
transformations
one expects the symmetry to be generated by some functional on phase
space, say $T^{\rm vir}(x)$, which would satisfy the (classical) Virasoro
algebra:
$$\left\{T^{\rm vir}(x),T^{\rm vir}(y)\right\}_2=(c/2)\delta^{\prime\prime
\prime}(x-y)-2T^{\rm vir}(x)\delta^\prime(x-y)-T^{{\rm vir}\prime}(x)
\delta(x-y).
$$
The component $L_{-1}=\int dx\,T^{\rm vir}(x)$
would generate translations in $x$, {\it i.e.\/}
$${\partial\phi(x)\over\partial x}=\left\{\phi(x),\int dy\,
T^{\rm vir}(y)\right\}_2.$$
When the centres are set to zero $x$ becomes identified with
$t_\Lambda$, and so such transformations are generated
by the flow associated with the Hamiltonian $H_\Lambda$. Therefore, up
to a possible total derivative, the Virasoro generator should be equal
to the Hamiltonian density $h^{-i}$, which one can readily confirm has scaling
dimension two, as required.

This leads us to the conclusion that the
second Poisson bracket algebra of a generalized KdV hierarchy
contains as a subalgebra the Virasoro
algebra, and is thus a (classical) chiral
extended conformal algebra.
We shall explicitly construct the Virasoro generator
for the examples that are considered in section 6.

\chapter{Modified Hierarchies and Miura Maps}

In this section we consider how our formalism
extends to the various {\it partially modified} hierarchies that are
associated to a given KdV hierarchy. These modified hierarchies are
constructed by considering a Lax operator of the form \Aa, where
$q\in C^\infty({\bf R}/{\bf Z},Q_{\bf s})$,
$Q_{\bf s}=\hat g_{\geq0}({\bf s})\bigcap\hat g^{<i}$.
The space  $Q_{\bf s}$
is a certain subspace of $Q$ labelled by another gradation of
$\hat g$, ${\bf s}_{\rm hom}\prec{\bf s}\preceq{\bf s}[w]$.
The fact that $Q_{\bf s}\subset Q$ is a consequence of
lemma 2.1 which implies that $Q=Q_{\bf s}\cup Y_{\bf s}$, where
$Y_{\bf s}=\hat g_0\cap\hat g_{<0}({\bf s})$. The modified
hierarchies have a gauge invariance of the form \Acc\ where $P$ is
replaced by $P_{\bf s}=\hat g_0({\bf s})\cap\hat g^{<0}$.
Thus the phase space, denoted ${\cal M}_{\bf s}$, of a partially
modified hierarchy consists of the
equivalence classes of operators of the form \Aa, with $q\in
C^\infty({\bf R}/{\bf Z},Q_{\bf s})$, modulo the gauge symmetry \Acc\
generated by $P_{\bf s}$.
Correspondingly, the space of gauge invariant
functionals on $Q_{\bf s}$ is denoted ${\cal F}_{\bf s}$.
The unique
hierarchy for which $P_{\bf s}=\emptyset$, {\it i.e.\/} ${\bf
s}\equiv{\bf s}[w]$ is the {\it modified\/} KdV hierarchy, whereas the
hierarchies for which ${\bf s}\prec{\bf s}[w]$ are known as {\it
partially modified\/} KdV hierarchies.
It was shown in ref. [\Rg] that
all the (partially) modified hierarchies associated to a KdV hierarchy
can be obtained as  reductions of that KdV hierarchy. Here we shall
prove this in a slightly different way through an analysis of the
Hamiltonian structure.

In attempting to extend the analysis of the previous
sections to the partially modified KdV hierarchies, we must separate the
results that explicitly require the first gradation ${\bf s}$ to be the
homogeneous gradation. An essential result in this direction is proposition
3.2, which demonstrates that the second Poisson bracket
can be defined without resorting to the homogeneous gradation.
{}From this result alone, we can predict that
the partially modified KdV hierarchies are
Hamiltonian with respect to the Poisson bracket
$$
\{\varphi,\psi\}_{\bf s}=\left(q_{\bf s}+\Lambda,\left[d_{q_{\bf s}}\varphi,
d_{q_{\bf s}}\psi\right]_{R_{\bf s}}\right)-\left(d_{q_{\bf s}}\varphi,\left
(d_{q_{\bf s}}\psi\right)^\prime\right),\eqn\PBprediction
$$
for $\varphi$ and $\psi$ $\in{\cal F}_{\bf s}$, $R_{\bf s}=(
P_{0[{\bf s}]}-P_{<0[{\bf s}]})/2$,
the Hamiltonians being defined identically to that of definition 3.1, with the
modification
that $q \in Q_{\bf s}$. Observe that the functional derivatives are valued in
$\hat g_{\leq 0}({\bf s})\big/ \hat g^{\leq-i}$.
This Hamiltonian property of the
pmKdV hierarchies can be
verified directly,
the dynamical equations of the partially modified KdV hierarchies
\Af, \Ag\ being reproduced, where subscripts now denote
the gradation ${\bf s}$. However, we derive this result through the
Hamiltonian mapping properties of the Miura map.

Recall that the phase space ${\cal M}_{\bf s}$ of a partially
modified hierarchy consists of the
equivalence classes of operators of the form \Aa, with $q\in
C^\infty({\bf R}/{\bf Z},Q_{\bf s})$, modulo the gauge symmetry \Acc\
generated by $P_{\bf s}$. Thus, if we define ${\rm Im}_{\bf s}(q)$ to
denote the $P_{\bf s}$-gauge orbit of $q$ in $Q_{\bf s}$,
the point of the phase space ${\cal M}_{\bf s}$
corresponding to $q \in C^\infty({\bf R}/{\bf Z},Q_{\bf s})$
is represented by the
gauge orbit Im$_{\bf s}(q)$.
Given the inclusion $Q_{\bf s} \subset Q$, there is a corresponding inclusion
of equivalence classes ${\cal M}_{\bf s} \subset {\cal M}$, where
the equivalence class
represented by ${\rm Im}_{\bf s}(q)$
is mapped into the equivalence class represented by ${\rm Im}_{{\rm
hom}}(q) \subset Q$.
This inclusion of
equivalence classes is the {\it Miura map} $\mu \colon\,
{\cal M}_{\bf s} \subset {\cal M}$, [\Sc].
Observe that the gauge group $P$ does not preserve
the subspace $Q_{\bf s}$, {\it i.e.\/} the image
${\rm Im}(\mu) \not\subset Q_{\bf s}$, where ${\rm Im}(\mu)=
\bigcup_{q \in Q_{{\bf s}}} {\rm Im}_{{\rm hom}}(q)$, the $P$-gauge
orbit of $Q_{\bf s}$ in $Q$.
There is an
induced mapping $\mu^* \colon\, {\cal F} \rightarrow {\cal F}_{\bf s}$
given by restriction of the functionals to the submanifold ${\cal
M}_{\bf s}$. In particular, we have the following proposition
relating the Hamiltonians.

\noindent
{\bf Proposition 5.1.}
{\it
The Hamiltonians of the
$(\Lambda,{\bf s},[w])$-partially modified hierarchy
are the restriction to ${\cal M}_s$ of the
Hamiltonians of the associated $(\Lambda,{\bf s_{\rm hom}},[w])$-KdV
hierarchy.}

\noindent
{\it Proof.}
This follows from the observation that by restricting
to the submanifold ${\cal M}_{\bf s}$,
we can perform a gauge transformation such that $q \in C^\infty({\bf
R}/{\bf Z},Q_{\bf s})$
in \Ae. Then $\Phi^j \equiv \Phi_{\bf s}^j \in C^\infty({\bf R}/{\bf
Z},\hat g_{\leq0}({\bf s})\cap\hat g^j)$,
{\it i.e.\/} $\Phi$ is identical to the unique transformation employed in the
$(\Lambda,{\bf s},[w])$-hierarchy.
Thus the restricted Hamiltonians in definition 3.1 are identical
to the Hamiltonians of the partially modified hierarchy.

Since the Hamiltonians of the partially modified hierarchy are reproduced
on restriction, the dynamical equations of the partially modified hierarchy
will be reproduced if the Hamiltonian structure restricts to
${\cal M}_{\bf s}$, {\it i.e.\/} if we define $I_{{\cal M}_{\bf s}}$
as the functionals that vanish on ${\cal M}_{\bf s} \subset {\cal M}$,
then a Hamiltonian structure is induced on
${\cal F}_{\bf s} \cong {\cal F} \big/ I_{{\cal M}_{\bf s}}$ if $I_{{\cal M}_
{\bf s}}$ is an
ideal of the Poisson bracket on ${\cal M}$.
This is proved in the following proposition:

\noindent
{\bf Proposition 5.2 }
{\it The second Hamiltonian structure
of the generalized KdV hierarchy induces the following Hamiltonian
structure on ${\cal M}_{\bf s}$ }:
$$
\{\varphi,\psi\}_{\bf s}=\left(q_{\bf s}+\Lambda,\left[d_{q_{\bf s}}\varphi,
d_{q_{\bf s}}\psi\right]_{R_{\bf s}}\right)-\left(d_{q_{\bf s}}\varphi,\left
(d_{q_{\bf s}}\psi\right)^\prime\right),\eqn\PBRs
$$
{\it for $\varphi$ and $\psi$ $\in{\cal F}_{\bf s}$,
$d_q\varphi \in C^\infty(
{\bf R}/{\bf Z},\hat g_{\leq 0}({\bf s})\big/
\hat g^{\leq-i})$,
and $R_{\bf s}=(
P_{0[{\bf s}]}-P_{<0[{\bf s}]})/2$.}

\noindent
{\it Proof.}
The idea of the proof is to verify that if we restrict
to ${\cal M}_{{\bf s}} \subset {\cal M}$, then the
corresponding operation on the functional derivatives, {\it i.e.\/}
taking the quotient, is well defined
as a Lie algebra homomorphism. The fact that the phase space
consists of gauge equivalence classes complicates this issue.
Thus we proceed as follows:
consider a functional $\varphi \in {\cal F}$.
For $q \in{\rm Im}(\mu)$, we can perform a gauge transformation
such that $q \in Q_{\bf s}$, the remaining gauge freedom being $P_{\bf
s}$. With this partial gauge fixing, the derivative of $\varphi$
with respect to the
directions $r\in Y_{\bf s}$ are not specified, {\it i.e.\/} the functional
derivative is
defined up to ${\rm Ann}(Q_{\bf s})=\hat g_0 \cap \hat
g_{>0}({\bf s})\ {\rm mod}\ \hat g^{\leq -i}$, where
${\rm Ann}(A)=\{l \in B^* \ | \left(l, a \right)=0\ \forall a \in A \}$
for $A$ a subalgebra of an algebra $B$.
Expressing the second Poisson
bracket in terms of the ${\bf s}$-gradation, proposition 3.2, we observe that
${\rm Ann}(Q_{\bf s})$ is an ideal of the centrally extended Lie algebra
$\{ \hat g_{\leq0}, ([\ ,\ ]_{R[{\bf s}]}, \omega_{\bf s}) \}$.
Thus the quotient is well defined, reproducing \PBRs.

Observe that this proves the prediction in \PBprediction, made on the
strength of proposition 3.2. The first Hamiltonian structure
has not been mentioned in relation to the partially modified hierarchies.
This is because although the first Poisson bracket is well defined on the
phase space ${\cal M}_{\bf s}$, it does not generate the dynamics, {\it
i.e\/}
equations \Af, \Ag\ are not reproduced.
{}From the Hamiltonian mapping point of view, this corresponds to
the fact that the space ${\rm Ann}(Q_{\bf s})$ is not an ideal
of the Lie bracket $[\ ,\ ]$ on $\hat g_{\leq0}$.

Combining these results, we obtain the following theorem.

\noindent
{\bf Theorem 5.1.}
{\it The Miura map is a Hamiltonian mapping\/},
$$\mu \colon\, \left({\cal M}_{\bf s},\{\ ,\ \}_{\bf s}\right)
\rightarrow \left({\cal M},\{\ ,\ \}_2\right),$$
{\it
such that it defines a reduction of the dynamical equations of
the KdV hierarchy to those of the pmKdV hierarchies.}

\noindent
{\it Proof.}
If we consider the decomposition $q=q_{\bf s}+\check q_{\bf s}$, where
$q \in C^\infty({\bf R}/{\bf Z},Q)$,
$q_{\bf s}\in C^\infty({\bf R}/{\bf Z},Q_{\bf s})$ and $\check q_{\bf s}
\in C^\infty({\bf R}/{\bf Z},Y_{\bf s})$, then
for a gauge equivalence class in ${\cal M}_{\bf s}$ the
Miura map is equivalent to the constraint $\check q_{\bf s}=0$.
The fact that the Miura map is Hamiltonian, as follows from proposition
5.2, implies that this constraint is preserved under all the flows.
Since the Hamiltonians are reproduced on restriction to
${\cal M}_{\bf s}$, the time evolutions of the pmKdV
hierarchies, \Af\ and \Ag, are reproduced under the Miura map.

We should emphasize that the Miura Map $\mu:\, {\cal M}_{\bf
s}\rightarrow{\cal M}$ is not invertible; it allows one to construct a
solution of the KdV hierarchy in terms of a solution of the
(partially) modified hierarchy, but not {\it vice-versa\/}. In addition,
one can show more generally that there exists a Miura Map between each
partially modified hierarchy $\mu:\ {\cal M}_{{\bf s}_1}\rightarrow
{\cal M}_{{\bf s}_2}$, whenever ${\bf s}_1\succ{\bf s}_2$.

\chapter{Examples}

\section{The Drinfel'd--Sokolov Generalized KdV Hierarchies}

The Drinfel'd-Sokolov KdV hierarchies [\Sc] are recovered from our formalism
by choosing $[w]$ to be the conjugacy class
containing the Coxeter element [\Rg]. In this case
$\Lambda=\sum_{i=1}^re_i+ze_0$, where the $e_i$, for $i=1,\ldots,r$
are the raising operators associated to the simple roots, and $e_0$ is
the lowering operator associated to the highest root. In this case
$Q=b_-$, one of the Borel subalgebras of $g$.
The gauge freedom corresponds to $n_-$, where $n_-$ is the
subalgebra of $g$ such that $b_-=n_-+h$ (as a vector space).
For example in the case
of $A_n$, choosing the defining representation, we have
$$\Lambda=\pmatrix{ &1&&\cr &&\ddots&\cr &&&1\cr z&&&\cr},\eqn\lamb$$
and $Q$ consists of the lower triangular matrices (including the
diagonal). In this example,
the gauge group is generated by the strictly lower
triangular matrices.

We can immediately write down the expression for the first and second
Hamiltonian structures. If we define $I\equiv\sum_{i=1}^re_i$ and
$e\equiv e_0$, to use the notation of [\Sc], then
$$\{\varphi,\psi\}_1=-\left(d_q\varphi,[d_q\psi,e]\right),$$
since $q$ has no $z$ dependence, and
$$\{\varphi,\psi\}_2=\left(d_q\varphi,[d_q\psi,\partial_x+q+I]\right).$$
These are exactly the two Hamiltonian structures of the generalized
KdV hierarchies written down in
[\Sc] for the untwisted Kac-Moody algebras. We have not considered the
case of twisted Kac-Moody algebras here, but it seems  that
in some cases only a single Hamiltonian structure exists, see [\Sc].

For $A_1$, one finds that using the basis for ${\cal F}$ in \bas\ one
recovers the explicit form of \kdv. The second Poisson bracket algebras
 are the Gel'fand-Dikii algebras [\Ri], which are
classical versions of the so-called $W$-algebras of ref. [\Rl]. For the
$A_n$ case, the scaling dimensions of the generators $u_j$ of \bas\ are
$n+2-j$, where
$j=1,\ldots,n$, and so the scaling dimensions range from $2$ to $n+1$ in
integer
steps. In this case ${\cal F}$ has a unique element of dimension $2$,
namely $u_n$, which generates the algebra of conformal transformations,
or the Virasoro algebra.

\section{The $A_1$ Fractional KdV--Hierarchies}

A series of hierarchies can be associated with any choice of $[w]$,
simply by choosing $\Lambda$ to be any element of ${\cal H}[w]$ with
well defined positive ${\bf s}[w]$-grade.
If we consider $g=A_1$ then there are two
elements in the Weyl group---the identity and the
reflection in the root. The identity leads to a homogeneous hierarchy
which is considered in section 6.5. Choosing $w$ to be the reflection,
the Heisenberg subalgebra is spanned by (in the defining
representation)
$$\Lambda^{2m+1}=z^m\pmatrix{0&1\cr z&0\cr},$$
where $m$ is an arbitrary integer, and the superscript denotes the
${\bf s}[w]$-grade. When one takes $\Lambda$ to be the ${\bf s}[w]$-grade
1 element, {\it i.e\/} $\Lambda^1$,
then the hierarchy which results is nothing but the usual
KdV hierarchy discussed above. When one takes $\Lambda$ to be an
element of the Heisenberg subalgebra with grade $>1$, then we have
what we might call a `fractional hierarchy', to mirror the terminology
of [\Se], because the fields $q$ have fractional scaling dimensions.

Let us consider these hierarchies in more detail. If we take
$\Lambda=\Lambda^{2m+1}$, {\it i.e.\/} $i=2m+1$, then before
gauge fixing, the potential is
$$
q=\sum_{j=0}^{m} z^j \pmatrix{a_{3j+2}&a_{3j+3}\cr
a_{3j+1}&-a_{3j+2}\cr},\quad\qquad a_{3m+3}=0.\eqn\PL
$$
The gauge transformation defined in \Ao\ involve the matrix
$$S=\pmatrix{1&0\cr A&1\cr},$$
and by choosing $A=a_{3m+2}$ one can generate a consistent gauge slice
$q^{\rm can}$ of the form \PL\ with $a_{3m+2}=0$, which
generalizes the
usual choice of canonical variables for the KdV hierarchy. For $m>0$, there
are $3m+1$ independent gauge invariant functionals, with $m$ of them,
corresponding to $h^{2k+1}$ for $k=0,\ldots,m-1$, in the center of the two
Poisson brackets. Rather than
present a general result for the two Hamiltonian structures, we just
consider the first two cases corresponding to $i=3$ and $i=5$; the
case $i=1$ is, of course, just the usual KdV hierarchy whose two
Poisson bracket algebras are written down in the introduction.

\subsection{The case $i=3$}

In this case there are 4 gauge invariant functionals. It is
straightforward to express them in terms of the variables $a_i$:
$$\eqalign{g_1&=a_1+2a_2a_5-a_5^2a_3-a_5^\prime\cr
g_2&=a_2-a_3a_5,\ \ \  g_3=a_3,\ \ \  g_4=a_3+a_4+a_5^2.\cr}$$
The $g_j$'s have scaling dimensions ${4\over3},1,{2\over3},{2\over3}$,
respectively. The non-zero brackets of the first symplectic structure are
$$\eqalign{\{g_2(x),g_1(y)\}_1&=(2g_3(x)-g_4(x))\delta(x-y)\cr
\{g_2(x),g_3(y)\}_1&=\delta(x-y),\ \ \
\{g_1(x),g_1(y)\}=2\delta^\prime(x-y).\cr}$$
In this case the variable $g_4$ is in the centre  as expected;
indeed, if the conserved densities are constructed one finds
$h^1=g_4$ . The non-zero brackets of the second structure are
$$\eqalign{\{g_1(x),g_3(y)\}_2&=-\delta^\prime(x-y)+2g_2(x)\delta(x-y)\cr
\{g_2(x),g_2(y)\}_2&=-{1\over2}\delta^\prime(x-y)\cr
\{g_2(x),g_1(y)\}_2&=g_1(x)\delta(x-y),\ \ \ \{g_2(x),g_3(y)\}_2=
-g_3(x)\delta(x-y).\cr}$$
Again, as expected from section 3.6, $g_4$ is in the centre of the
algebra. We recognize the algebra as the $A_1$ Kac-Moody algebra with
non-trivial central extension. It is straightforward to write down the
Virasoro generator for this algebra
$$T^{\rm vir}(x)=g_1(x)g_3(x)+g_2(x)^2 -{1\over3}g'_{2}(x).$$

\subsection{The case $i=5$}

The space of gauge invariant functions is spanned by integrals of
polynomials in the seven gauge invariant functions
$$\eqalign{g_1&=a_1+2a_2a_8-a_8^2a_3-a_8^\prime\cr
g_2&=a_2-a_8a_3,\ \ \ g_3=a_3\cr
g_4&=a_3+a_4+2a_5a_8+a_6a_7\cr
g_5&=a_5-a_6a_8\cr
g_6&=a_6,\ \ \ g_7=a_6+a_7+a_8^2.\cr}$$
The spins of the functions $g_j$ are
${6\over5},1,{4\over5},{4\over5},{3\over5},{2\over5},{2\over5}$,
respectively. The non-zero brackets
of the first Poisson bracket algebra are
$$\eqalign{\{g_2(x),g_1(y)\}_1&=(2g_3(x)-g_4(x)-g^{2}_{6}(x)+g_6(x)g_7(x))
\delta(x-y)\cr
\{g_2(x),g_3(y)\}_1&=g_6(x)\delta(x-y),\ \ \
\{g_2(x),g_6(y)\}_1=\delta(x-y)\cr
\{g_1(x),g_1(y)\}_1&=2\delta^\prime(x-y),\ \ \ \{g_1(x),g_3(y)\}_1=-2g_5
(x)\delta(x-y)\cr \{g_1(x),g_5(y)\}_1&=(g_7(x)-2g_6(x))\delta(x-y)\cr
\{g_3(x),g_5(y)\}_1&=-\delta(x-y).\cr}$$
There are two centres $g_4$ and $g_7$ as predicted by proposition 3.3.
The non-zero brackets of the second Poisson bracket algebra are
$$\eqalign{\{g_2(x),g_2(y)\}_2&=-{1\over2}\delta^\prime(x-y),\ \ \
\{g_2(x),g_1(y)\}_2=g_1(x)\delta(x-y)\cr
\{g_2(x),g_3(y)\}_2&=-g_3(x)\delta(x-y),\cr
\{g_1(x),g_3(y)\}_2&=-\delta^\prime(x-y)+2g_2(x)\delta(x-y)\cr
\{g_5(x),g_6(y)\}_2&=\delta(x-y).\cr}$$
Again $g_4$ and $g_7$ are centres of the algebra. The second Poisson
bracket algebra is simply the direct sum of an $A_1$ Kac-Moody
algebra, with central extension, generated by $g_1$, $g_2$ and $g_3$,
and a `$b$-$c$' algebra, generated by $g_5$ and $g_6$. The Virasoro
generator is, in this case, given by
$$T^{\rm vir}(x)=g_1(x)g_3(x)+g_2(x)^2-{1\over3}g'_2(x)+g_5(x)g_6^\prime(x)
-{2\over5}(g_5(x)g_6(x))'.$$

\section{The First Fractional $A_2$ KdV--Hierarchy and $W_3^{(2)}$}

Let us consider the KdV hierarchy corresponding to the Coxeter element
of the Weyl group, but in contrast to the usual Drinfel'd-Sokolov
case, where $\Lambda$ is given by \lamb, we now
take $\Lambda$ to be the element of the Heisenberg subalgebra with
$i=2$, {\it i.e.\/}
$$\Lambda=\pmatrix{0&0&1\cr z&0&0\cr 0&z&0\cr}.$$
This choice corresponds to the fractional KdV hierarchy discussed in [\Rg,\Se].
Before gauge fixing the potential can be written as
$$
q=\pmatrix{y_1&c&0\cr
           e&y_2&d\cr
           a+bz&f&-(y_2+y_1)\cr},
$$
which, under a gauge transformation, transforms as
$$
q\rightarrow \tilde q = \Phi\partial_x\Phi^{-1} + \Phi(q+\Lambda)\Phi^{-1} -
\Lambda,
$$
where
$$
\Phi=\pmatrix{1&0&0\cr
              A&1&0\cr
              B&C&1\cr}.
$$
As shown in ref. [\Rg]
there exists a gauge transformation given by
$$
\eqalign{
A=& {1\over3}(b+c-2d)\cr
C=&{1\over3}(2c-b-d)\cr
B=&y_1+y_2 - dC -{\beta\over\alpha}(cA+y_2-dC-AC),\cr}
$$
which brings $q$ into the canonical form
$$
q^{\rm can}=\pmatrix{(\alpha-\beta)U&0&0\cr
                  G^+&-\alpha U&0\cr
                  T&G^-&\beta U\cr}
+ \phi \pmatrix{0&1&0\cr 0&0&1 \cr z&0&0\cr},
$$
where $\alpha$, $\beta$ are arbitrary parameters which we fix to
$\alpha=2\beta=1$, to make the comparison
with the results of ref. [\Se] easier. The fields $\phi$, $U$,
$G^{\pm}$ and $T$ form a basis of gauge invariant functionals of $q$, with
spins ${1\over2}$, $1$, ${3\over2}$ and $2$ . The field $\phi$ corresponds
to the conserved
quantity $h^{1}$ which, following section 3.6, is in the center of the two
Poisson brackets.

In terms of the combinations
$$
\eqalign{
\tilde U&= U +\phi^2, \quad  \qquad \tilde G^{\pm} = G^{\pm} \pm \phi'
- {3\over2}U\phi -\phi^3\cr
\tilde T& =T + {3\over4}U^2 + (G^+ +G^-)\phi,\cr}
$$
the only non-vanishing Poisson
brackets in the first Hamiltonian structure read
$$
\eqalign{
\{\tilde U(x),\tilde G^{\pm}(y)\}_1& = \mp \delta(x-y) \cr
\{\tilde G^+(x),\tilde G^-(y)\}_1 &=-3\phi(x) \delta(x-y)\cr
\{\tilde T(x),\tilde G^{\pm}(y)\}_1& = {3\over2}\delta^\prime(x-y)\cr
\{\tilde T(x),\tilde T(y)\}_1 & = 6\phi(x) \delta^\prime(x-y)+3\phi'(x)
\delta(x-y),\cr}
$$
while, in the second structure, the non-vanishing brackets are
$$
\eqalign{
\{\tilde U(x),\tilde U(y)\}_2 &= -{2\over3}\delta^\prime(x-y)\cr
\{\tilde U(x),\tilde G^{\pm}(y)\}_2 & = \pm\tilde G^{\pm}(x)\delta(x-y) \cr
\{\tilde G^+(x),\tilde G^-(y)\}_2& = -\delta^{\prime\prime}(x-y)
+ 3 \tilde U(x) \delta^\prime(x-y)\cr &\ \ \ \ \ \ \
+ \left(\tilde T(x) + {3\over2} \tilde U'(x) - 3\tilde U^2(x)\right)
\delta (x-y)\cr
\{\tilde T(x),\tilde U(y)\}_2 & = -\tilde U(x)\delta^\prime(x-y) \cr
\{\tilde T(x) ,\tilde G^{\pm}(y)\}_2& =-{3\over2} \tilde G^{\pm}(x)
\delta^\prime(x-y)
- {1\over2}\tilde G'^{\pm}(x) \delta(x-y)\cr
\{\tilde T(x),\tilde T(y)\}_2& =
{1\over2}\delta^{\prime\prime\prime}(x-y) -
2\tilde T(x)\delta^\prime(x-y) - \tilde T'(x)\delta(x-y). \cr}
$$
As explained before, the second Hamiltonian structure is an extension of the
Virasoro algebra which, in this case, corresponds to the generalized
$W$-algebra $W^{(2)}_{3}$ [\Sg], in agreement with ref. [\Se], where
$\tilde T\equiv T^{\rm vir}$ is the Virasoro generator.

For simplicity, we have only considered the case of $W_3^{(2)}$ here;
however,
our construction can easily be extended to more complicated cases, the
complexity of the equations being the only obstacle to such an
endeavor.

\section{The Hierarchy Associated to $w=R_{\alpha_0}$ in $A_2$}

The Weyl group of $A_2$ has three conjugacy classes. One contains the
Coxeter element, which leads to the Drinfel'd-Sokolov KdV hierarchies and
their fractional generalizations considered above. The identity element of
the Weyl group leads to a {\it homogeneous hierarchy\/}, which are
considered
below. In this section, we consider the third possibility.
We take as our representative of the conjugacy class
the reflection in the root
$\alpha_0 = -\alpha_1 -\alpha_2$, where $\alpha_1$ and $\alpha_2$ are the
simple roots. For a description of how the Heisenberg subalgebra
is constructed in this case, we refer to [\Rg]. The
simplest KdV hierarchy associated to this conjugacy class
is obtained by taking $\Lambda$
to be the element of the Heisenberg subalgebra with lowest grade, in this
case $i=2$, {\it i.e.\/}
$$
\Lambda=\Lambda_{2,0}=\pmatrix{0&0&1\cr
                 0&0&0\cr
                 z&0&0\cr}.
$$
Following ref. [\Rg], the potential, before gauge fixing, can be written as
$$
q=\pmatrix{y_1&c&0\cr
           e&y_2&d\cr
           a&f&-(y_2+y_1)\cr},
$$
which under a gauge transformation changes to
$$
q\rightarrow \tilde q = \Phi\partial_x\Phi^{-1} + \Phi(q+\Lambda)\Phi^{-1} -
\Lambda,
$$
where
$$
\Phi=\pmatrix{1&0&0\cr
              A&1&0\cr
              B&C&1\cr}.
$$

In this case, there exists a gauge transformation given by
$$
A= -d,\qquad B=y_1 +{1\over2}(y_2 - cd), \qquad C=c,
$$
which brings $q$ into the canonical form
$$
q^{\rm can}=\pmatrix{0&0&0\cr
                 G^+&0&0\cr
                 T&G^-&0\cr}
+ {U\over 2} \pmatrix{1&0&0\cr
                      0&-2&0\cr
                      0&0&1\cr}.
$$
Again, $U$, $G^{\pm}$ and $T$ are gauge invariant functionals of $q$.
Notice that
$U$ corresponds to the conserved quantity $h^0$, which,
following section 3.6, is only in
the centre of the first Poisson bracket and not the second.

The  non-vanishing Poisson brackets are the following. For the first
Hamiltonian structure they are
$$
\eqalign{
\{G^+(x),G^-(y)\}_1& =\delta(x-y)\cr
\{T(x),T(y)\}_1 & = -2 \partial_x\delta(x-y), \cr}
$$
while for the second structure they are
$$
\eqalign{
\{U(x),U(y)\}_2  &= -{2\over3}\delta^\prime(x-y)\cr
\{U(x),G^{\pm}(y)\}_2 & = \pm G^{\pm}(x)\delta(x-y) \cr
\{G^+(x),G^-(y)\}_2& = -\delta^{\prime\prime}(x-y)
+ 3 U(x)\delta^\prime(x-y)\cr
&\ \ \ \ \ \ +\left(T(x) + {3\over2} U'(x) - 3 U^2(x)
\right) \delta (x-y)\cr
\{T(x),U(y)\}_2 & = -U(x)\delta^\prime(x-y) \cr
\{T(x),G^{\pm}(y)\}_2& =-{3\over2} G^{\pm}(x)
\delta^\prime(x-y)- {1\over2}G'^{\pm}(x)\delta(x-y)  \cr
\{T(x),T(y)\}_2& = {1\over2}\delta^{\prime\prime\prime}(x-y) -
2T(x)\delta^\prime(x-y) - T'(x)\delta(x-y).\cr}
$$

In this case, the extension of the Virasoro algebra described by the second
Poisson bracket is again the generalized W--algebra $W_{3}^{(2)}$, with
$T\equiv T^{\rm vir}$
being the Virasoro generator. That the same algebra should appear in this
example and in that of 6.3 can clearly be seen  from the
definitions of the brackets.
Nevertheless,  even though the second Hamiltonian structures are identical,
 the two
hierarchies of partial differential equations are completely different.

\section{The Homogeneous Hierarchies}

The homogeneous hierarchies were defined in [\Rg]. They arise from
taking ${\bf s}[w]$ to be the homogeneous gradation, corresponding to
the identity element of the Weyl group. The simplest such hierarchy has
$\Lambda=z\mu\cdot H$ and $q=f\cdot H+\sum_{\alpha\in\Phi_g}q^\alpha
E_\alpha$, where $\{H,E_\alpha\ \alpha\in\Phi_g\}$ is a Cartan-Weyl
basis for $g$, and $f$ and $q^\alpha$ are the dynamical variables. In
order that the hierarchy be of type I, $\mu\cdot H$ must be regular,
which implies that $\mu\cdot\alpha\neq0$ $\forall\alpha\in\Phi_g$. It was
observed in [\Rg] that $h^0=f\cdot H$, for this hierarchy, and so the
variables $f$ are constant for all the flows.

The first and second symplectic structures are easily calculated for
the example to hand.
In the first case one finds that the non-zero brackets are
$$\{q^\alpha(x),q^\beta(y)\}_1=(\mu\cdot\alpha)\delta_{\alpha+\beta,0}
\delta(x-y),$$
and so the variables $f$ are indeed centres as proved in proposition
3.6. In the second case the non-zero brackets are
$$\eqalign{\{\nu\cdot f(x),\lambda\cdot f(y)\}_2&=(\nu\cdot\lambda)\delta
^\prime(x-y)\cr
\{q^\alpha(x),\nu\cdot f(y)\}_2&=(\alpha\cdot\nu)q^\alpha(x)\delta(x-y)\cr
\{q^\alpha(x),q^\beta(y)\}_2&=\delta_{\alpha+\beta,0}\left(\delta^\prime
(x-y)-\alpha\cdot f(x)\delta(x-y)\right)\cr &\ \ \ \ \ \
+\epsilon(-\alpha,-\beta)q^{\alpha+\beta}(x)\delta(x-y),\cr}$$
where $\epsilon(\alpha,\beta)$ is non-zero if $\alpha+\beta$ is a root of $g$.
 So the
second symplectic structure is nothing but the Kac-Moody algebra $\hat
g$ with a central extension. Notice that $f$ is not in the centre of the
second Poisson bracket algebra, an eventuality previously encountered in
proposition 3.3.
Nevertheless, $f$ is a constant under the flows of the hierarchy
because the Hamiltonians satisfy the functional equation
$$\{f(x),H\}_2=0.$$
The Virasoro generator, in this case, is constructed from the
fields $f$ and $q^\alpha$ via the
Sugawara construction.

\chapter{Discussion}

We have presented a systematic discussion of the Hamiltonian structure
of the hierarchies of integrable partial differential equations constructed in
ref. [\Rg]. It was found that the analogues of the
KdV hierarchies admit two distinct yet coordinated Hamiltonian
structures, whereas the associated partially modified hierarchies only admit a
single Hamiltonian structure, generalizing the results of Drinfel'd
and Sokolov.
In addition, we found that
the Miura map between a modified hierarchy and its
associated KdV hierarchy, with the second
Hamiltonian structure, is a Hamiltonian map.
An aspect of the analysis that we have ignored is a thorough discussion of
the restriction of the phase space $\cal M$ to a symplectic leaf, {\it
i.e.\/} the r\^ole of the centres.
This will be considered in a later publication, where we
propose a group theoretic description of these hierarchies in terms of
the AKS/coadjoint formulation of integrable systems.
Ultimately it would be desirable to
understand the relation between the Poisson brackets
on the phase space ${\cal M}$ presented here, and
the Poisson brackets that exist on
the Akhiezer-Baker functions [\Rshansky].
If the analysis in [\Rshansky] generalises to the continuum
limit, this would lead to an
interpretation of the {\it dressing transformation\/} in terms of a
Hamiltonian mapping. Connected with this,
we are also intrigued by the relation of
our work to that of V.G. Kac and M. Wakimoto [\Re], who, following the
philosophy of the Japanese school, construct a hierarchy associated
to each of the basic (level 1) representations of a Kac-Moody algebra,
and each conjugacy class of the Weyl group of the underlying finite
Lie algebra. These
hierarchies are intimately connected to the vertex operator representations
of Kac-Moody algebras; see [\Re] for further details.

Note that only the untwisted Kac-Moody
algebras have been considered in this paper;
it appears that the KdV hierarchies associated to twisted Kac-Moody
algebras sometimes only admit a single Hamiltonian structure, see [\Sc].

The Second Hamiltonian structure is invariant under an arbitrary
conformal transformation. These transformations include the scale
transformations which reflect the quasi-homogeneity of the equations of the
hierarchy, that is to say all quantities have well defined scaling dimensions
such that a scale transformation leaves the equations of motion invariant.
On general principles, one would expect that the second
Poisson bracket algebra should contain the (chiral) algebra of conformal
transformations as a subalgebra, {\it i.e.\/} the Virasoro algebra.
This would imply that the second Poisson
bracket algebra is an extended (chiral) conformal algebra, generalizing the
appearance of the $W_n$ algebras in the work of Drinfel'd and Sokolov
[\Sc] and Gel'fand and Dikii [\Ri]. This was indeed found to be the case for
the
examples that were considered.

Of particular interest is the question as to whether these hierarchies have
any r\^ole to play in the non-perturbative structure of two dimensional gravity
coupled to matter systems, generalizing the known connexion of the
Drinfel'd-Sokolov hierarchies. It seems that one must supplement the hierarchy
with an additional equation, the so-called {\it string equation\/}, which
has to be
consistent with the flows of the hierarchy. Then  the potentials of the
hierarchy are apparently
related to certain correlation functions of the field theory.
Details of this will be presented elsewhere.

\ack
We would like to thank E. Witten for motivating our
investigation of the integrable hierarchies of KdV-type.
The research reported in this paper was performed under
the following grants:
The research of JLM is supported by a
Fullbright/MEC fellowship, that of MFdeG by the Natural
Sciences and Engineering Research Council of Canada, that of TJH
by NSF\# PHY 90-21984 and that of NJB by NSF\# PHY 86-20266.

While preparing this preprint we received ref. [\Rm], in which the Hamiltonian
structures of the $W_3^{(2)}$ algebra are discussed.

\par \penalty-400 \vskip\chapterskip
    \spacecheck\referenceminspace \immediate\closeout\referencewrite
    \referenceopenfalse
    \line{\fourteenrm\hfil REFERENCES\hfil}\vskip\headskip
    \input referenc.texauxil

\end